\documentclass[preprint]{aastex63}

\newcommand{\sgr}{\mbox{SGR\,J1935+2154~}}
\newcommand{\sgrnos}{\mbox{SGR\,J1935+2154}}

\received{\today}
\revised{\today}
\accepted{\today}

\submitjournal{ApJ}

\shorttitle{\sgr bursts}
\shortauthors{Lin et al.}

\graphicspath{{./}{figures/}}

\begin{document}

\title{Burst properties of the most recurring transient magnetar \sgrnos}

\correspondingauthor{Lin Lin}
\email{llin@bnu.edu.cn}

\author[0000-0002-0633-5325]{Lin Lin}
\affiliation{Department of Astronomy, Beijing Normal University, Beijing 100875, China}

\author{Ersin G\"o\u{g}\"u\c{s}}
\affiliation{Sabanc\i~University, Faculty of Engineering and Natural
Sciences, \.Istanbul 34956 Turkey}

\author{Oliver J. Roberts}
\affiliation{Universities Space Research Association, 320 Sparkman Drive, Huntsville, AL 35805, USA}

\author{Chryssa Kouveliotou}
\affiliation{Department of Physics, The George Washington University, 725 21st Street 
NW, Washington, DC 20052, USA}
\affiliation{Astronomy, Physics, and Statistics Institute of Sciences (APSIS), The 
George Washington University, Washington, DC 20052, USA}

\author{Yuki Kaneko}
\affiliation{Sabanc\i~University, Faculty of Engineering and Natural
Sciences, \.Istanbul 34956 Turkey}

\author{Alexander J. van der Horst}
\affiliation{Department of Physics, The George Washington University, 725 21st Street 
NW, Washington, DC 20052, USA}
\affiliation{Astronomy, Physics, and Statistics Institute of Sciences (APSIS), The 
George Washington University, Washington, DC 20052, USA}

\author{George Younes}
\affiliation{Department of Physics, The George Washington University, 725 21st Street 
NW, Washington, DC 20052, USA}
\affiliation{Astronomy, Physics, and Statistics Institute of Sciences (APSIS), The 
George Washington University, Washington, DC 20052, USA}

\begin{abstract}

We present timing and time-integrated spectral analysis of 127 bursts from \sgrnos. These bursts were observed with the Gamma-ray Burst Monitor on the \textit{Fermi Gamma-ray Space Telescope} and the Burst Alert Telescope on the \textit{Neil Gehrels Swift Observatory} during the source's four active episodes from 2014 to 2016. This activation frequency makes \sgr  the most burst prolific transient magnetar. We find the average duration of all the detected bursts to be much shorter than the typical, anticipated value. We fit the burst time-integrated spectra with two black-body functions, a Comptonized model and three other simpler models. Bursts from \sgr exhibit similar spectral properties to other magnetars, with the exception of the power law index from the Comptonized model, which correlates with burst fluence. We find that the durations and both black-body temperatures of the bursts have significantly evolved across the four active episodes. We also find that the burst time history exhibits two trends, which are strongly correlated with the decay of the persistent emission in each outburst. 

\end{abstract}

\keywords{soft gamma repeater: general --- soft gamma repeater: individual (\sgrnos)}

\section{Introduction} \label{sec:intro}
Magnetars comprise a group of isolated neutron stars with extremely strong magnetic fields \citep{kouveliotou1998}, which fully determine the emission properties of these systems. Magnetars slow down rapidly ($\dot{P}\thicksim 10^{-13}-10^{-11}~\rm{s~s}^{-1}$) likely under the influence of large magnetic torques. As a result, their rotational periods are very slow ($P \thicksim 2-12~$ s), even though they are young objects, typically 10$^3$ years old. Recently, magnetar-like activity was detected from the central compact object of supernova remnant, RCS~103. With a rotational period of $6.67$~hrs, it may be the longest of any observed magnetar to-date \citep{rea2016,dai2016}. The typical X-ray luminosity of a magnetar ranges between $10^{33}-10^{36}~\rm{erg~s}^{-1}$, which exceeds their rotational energy losses by a few orders of magnitude. Consequently, it has been suggested that the X-ray emission in magnetars is powered by the decay of their extreme magnetic fields ($B\sim 10^{14}-10^{15}~\rm{G}$). However, two sources were discovered in the last decade with relatively low inferred dipole fields, of the order of  $B\sim 10^{12}\, G$ \citep{rea2010, rea2012, rea2014, an2019}. The spectral properties of these two sources suggest that their surface magnetic field is still comparable to those of the magnetar population, albeit, with a more complicated configuration than a simple dipole field \citep{guver2011, tiengo2013}.

The majority of magnetars undergo occasional random outbursts during which time, their persistent emission increases significantly while simultaneously emitting bursts (or intermediate flares), in the hard X-ray or soft $\gamma-$ray energy regime. So far, we have detected bursts from 18 out of 23 confirmed magnetars \citep{olausen2014}. Based on their duration and peak luminosities, magnetar bursts can be classified into three types. Giant Flares, which are the rarest and most energetic magnetar events. Only three of this kind have been observed, from three different sources \citep{mazets1979, hurley1999, hurley2005, palmer2005}. All three events start with an initial hard peak lasting $0.1-0.2~\rm{s}$ with a luminosity of $10^{44}-10^{47}~\rm{erg~s}^{-1}$, followed by a spin-period-modulated soft tail, lasting hundreds of seconds. Intermediate flares last $1-40~\rm{s}$ and have peak luminosities of $10^{41}-10^{43}~\rm{erg~s}^{-1}$. The most common events are short bursts, with a typical duration of $\sim0.1~\rm{s}$ and peak luminosities of $10^{39}-10^{41}~\rm{erg~s}^{-1}$. Bursts and flares are also powered by the magnetic field, either through neutron star crustquakes \citep{td1995} or via magnetic field line reconnection \citep{lyutikov2003}. For a recent magnetar burst review, see \citet{turolla2015}.

During an outburst, the persistent X-ray luminosity of the source may increase at a rate $10-1000~$ times its quiescent level. The flux then gradually decays back to a pre-outburst level on timescales from weeks to years \citep{rea2011,coti2018}. Almost all magnetar observed outbursts are accompanied by the detection of bursting activity. However, there is no unique source activation trend. For example, Magnetar $1$E$~1841-045$ showed no obvious change in its persistent X-ray emission after emitting short bursts \citep{lin2011b}. Conversely, there are transient magnetars that have remarkable increases in their persistent X-ray flux after their bursting episode(s). 

\sgr was discovered after emitting a short burst which triggered the Burst Alert Telescope (BAT) aboard the \textit{Neil Gehrels Swift Observatory} (hereafter \textit{Swift}), on 2014 July 5. Follow-up observations carried out between July, 2014 and March, 2015 with \textit{Chandra} and \textit{XMM-Newton} allowed the measurement of its spin period and spin-down rate, found to be $P=3.24~\rm{s}$ and $\dot P= 1.43(1)\times10^{-11}~\rm{s~s}^{-1}$, respectively. This implies a dipole-magnetic field of $B\sim2.2\times10^{14}~\rm{G}$ \citep{israel2016}, confirming its magnetar nature. Since its discovery, \sgr exhibited burst active episodes almost annually, becoming the most recurring transient magnetar ever observed. \citet{kozlova2016} reported an intermediate flare from \sgr detected by four Interplanetary network (IPN) spacecraft on April 12$^{th}$ 2015 (not observed by the Gamma-ray Burst Monitor (GBM) aboard \textit{Fermi}). The flare lasted for $\sim1.7~\rm{s}$ with an energy fluence of $\sim2.5\times10^{-5}~\rm{erg~cm}^{-2}$. The source went into outburst in 2014 \citep{israel2016}, 2015, and twice in 2016.  During the 2015 outburst, a hard X-ray spectral component was revealed in the persistent source spectrum with \textit{NuSTAR} observations. During the 2016 outbursts, the soft X-ray flux was found to have increased by about 7 times the previous reported levels \citep{younes2017}.

We report the results of our extensive search for short bursts from this prolific transient magnetar, using a Bayesian block method to search the \textit{Swift}/BAT and \textit{Fermi}/GBM data. We performed detailed temporal and spectral analyses on all identified short bursts to establish not only their collective statistical properties, but any characteristic variations during each burst-active episode. The layout of our study is as follows: The data reduction and burst search procedure are introduced in Section \ref{sec:obs}. In Section \ref{sec:result}, we present the detailed spectral and temporal analyses for all reported bursts, including their lightcurves and localization. We discuss our results in Section \ref{sec:dis}.

\section{Observations and burst sample} \label{sec:obs}

The \textit{Swift}/BAT is a sensitive, mask-coded imaging instrument with a 1.4 steradian partially coded field of view (FoV) in the $15-150~\rm{keV}$ energy band \citep{barthelmy2005}. Nominally, BAT works in a surveying mode, which only provides detector plan histograms integrated for about five minutes. When the instrument is triggered by a burst, it shifts to a “burst mode”, and the time tagged event list is restored. This event list covers a time interval of about $-200$ to 2000 s either side of the trigger time, with a time resolution of $\sim0.2~\rm{ms}$. Considering that most magnetar bursts are short, we only study the time-tagged event list data.  BAT was triggered 10 times by \sgr bursts between 2014 and 2016. Their observational IDs are listed in Table \ref{tab:batobs}.

\begin{deluxetable}{ccc}
\tablenum{1}
\tablecaption{\textit{Swift}/BAT \sgr bursts. \label{tab:batobs}}
\tablewidth{0pt}
\tablehead{
\colhead{Observational ID } & \colhead{Date} & \colhead{Number of Bursts}
}
\startdata
$00603488000$ & 2014-07-05 & 3 \\
$00632158000$ & 2015-02-22 & 1 \\
$00632159000$ & 2015-02-22 & 1 \\
$00686443000$ & 2016-05-16 & 1 \\
$00686761000$ & 2016-05-18 & 2 \\
$00686842000$ & 2016-05-19 & 1 \\
$00687123000$ & 2016-05-21 & 3 \\
$00687124000$ & 2016-05-21 & 3 \\
$00701182000$ & 2016-06-23 & 4 \\
$00701590000$ & 2016-06-26 & 2 \\
\enddata
\end{deluxetable}

The \textit{Fermi}/GBM comprises 12 NaI(Tl) detectors ($\sim8$ keV $-$ 1 Mev), each with a diameter of 12.7~cm and length of 1.27~cm. The detectors are located in clusters of three at each of the four corners of the spacecraft \citep{meegan2009}. GBM also has two BGO detectors on opposing sides of the spacecraft, however they are not used in this analysis as the spectral range of the magnetars reported in this study lie below their effective energy range (0.2$-$40 MeV). As an all-sky monitor, GBM has an unocculted FoV of 8 steradians. Since 2012, GBM data are recorded in the continuous time-tagged event (TTE) mode with a fine temporal resolution of $2~\mu\rm{s}$. The spectral resolution comprises 128 pseudo-logarithmically scaled channels over an energy range of $8-1000~\rm{keV}$. These data types and its monitoring nature make GBM ideal for Magnetar burst studies. GBM was triggered 62 times by bursts from \sgr from $2014-2016$.

Not all Magnetar bursts triggered both BAT and GBM, even when the source was within the FoV of both instruments. Due to the triggering settings of GBM, new bursts that occur five minutes after a previous trigger, or weak bursts below the trigger thresholds, are usually not picked up. In addition to GBM being unable to trigger five minutes following a burst trigger, the thresholds on the BAT are increased for re-triggering on a known source. An un-triggered search for bursts through the entire available data is therefore essential in order to have a burst history that is as complete as possible for any magnetar source.  

The Bayesian block method is a non-parametric modeling technique for detecting and characterizing local variability in time-series data \citep{scargle2013}. The automated process divides the binned lightcurve or time-tagged event list into blocks, each block being consistent with a constant rate. It then finds the optimal segmentation or boundaries between the blocks by maximizing likelihood, termed “change points.” These step functions have no priors in amplitude nor duration. This method has been used in standard BAT data analysis procedures to calculate the duration of Gamma-Ray Bursts (GRBs), especially in helping identify the extended emission following some short GRBs \citep{norris2010,kaneko2015}. \citet{lin2013} applied this method to identify weak bursts in \textit{XMM-Newton} and \textit{Swift}/X-ray Telescope (XRT) observations of two magnetars and found the properties of those bursts to be dimmer by $1-2$ orders of magnitude than the triggered ones.

We performed the un-triggered burst search using the Bayesian block method on the BAT triggered-event data and GBM continuous TTE data. For the BAT data, we first extracted the mask-weighted lightcurve over the $15-150~\rm{keV}$ band with $4~\rm{ms}$ temporal resolution for each trigger. We then searched through the lightcurves using the same two-step procedure described in \citet{gogus2016}. We found 11 additional events in the 10 BAT triggered event data sets. The number of bursts found in each observation is listed in Table \ref{tab:batobs}. For each BAT burst we extracted the spectrum using \textit{batbinevt}, made required corrections (\textit{batupdatephakw} and \textit{batphasyserr}), and generated responses (\textit{batdrmgen}).  We then performed time-integrated spectral analysis with variable bin-sizes of at least $1\sigma$ significance, using \textit{XSpec} \citep{arnaud1996} and $\chi^2$ statistics.  

We also performed the Bayesian block search over the continuous TTE GBM data for the following time intervals: July 1$^{st}$ -- 15$^{th}$, 2014; February 15$^{th}$ -- April 15$^{th}$, 2015; and January 1$^{st}$ -- October 31$^{st}$, 2016. The search procedure is similar to what we used in \citet{lin2013}, with some modifications to the parameters. In order to limit computation time, we rebinned the TTE data over an energy range of $10-100~\rm{keV}$ into $8~\rm{ms}$. We then started a two-round search using a timing window of $8~\rm{s}$. We repeated the same search for all 12 NaI(Tl) detectors, flagging simultaneous events detected in two or more detectors. Detectors with an angle to the source of less than $60^{\circ}$\footnote{Results do not change significantly when choosing a smaller angle, e.g., $40^{\circ}$}, without any blockage by the satellite were then chosen and their location was calculated on the sky. We found 112 \sgr bursts in the GBM data, including 62 triggered events. Overall, there are 127 unique bursts from \sgr observed with BAT and GBM, with six events simultaneously recorded by both instruments. The ID, instrument information, and burst start time for all 127 bursts are listed in Table \ref{tab:burstlist}. We performed a spectral fit to each of these bursts with the standard GBM analysis software \textit{RMFIT} using Castor C-statistics (\textit{c-stat}). The detector response matrices were generated with \textit{GBMDRM v2.0}. 

\section{Results} \label{sec:result}

\subsection{Burst activity history}

\begin{figure*}
\includegraphics*[viewport=75 160 600 495, scale=0.5]{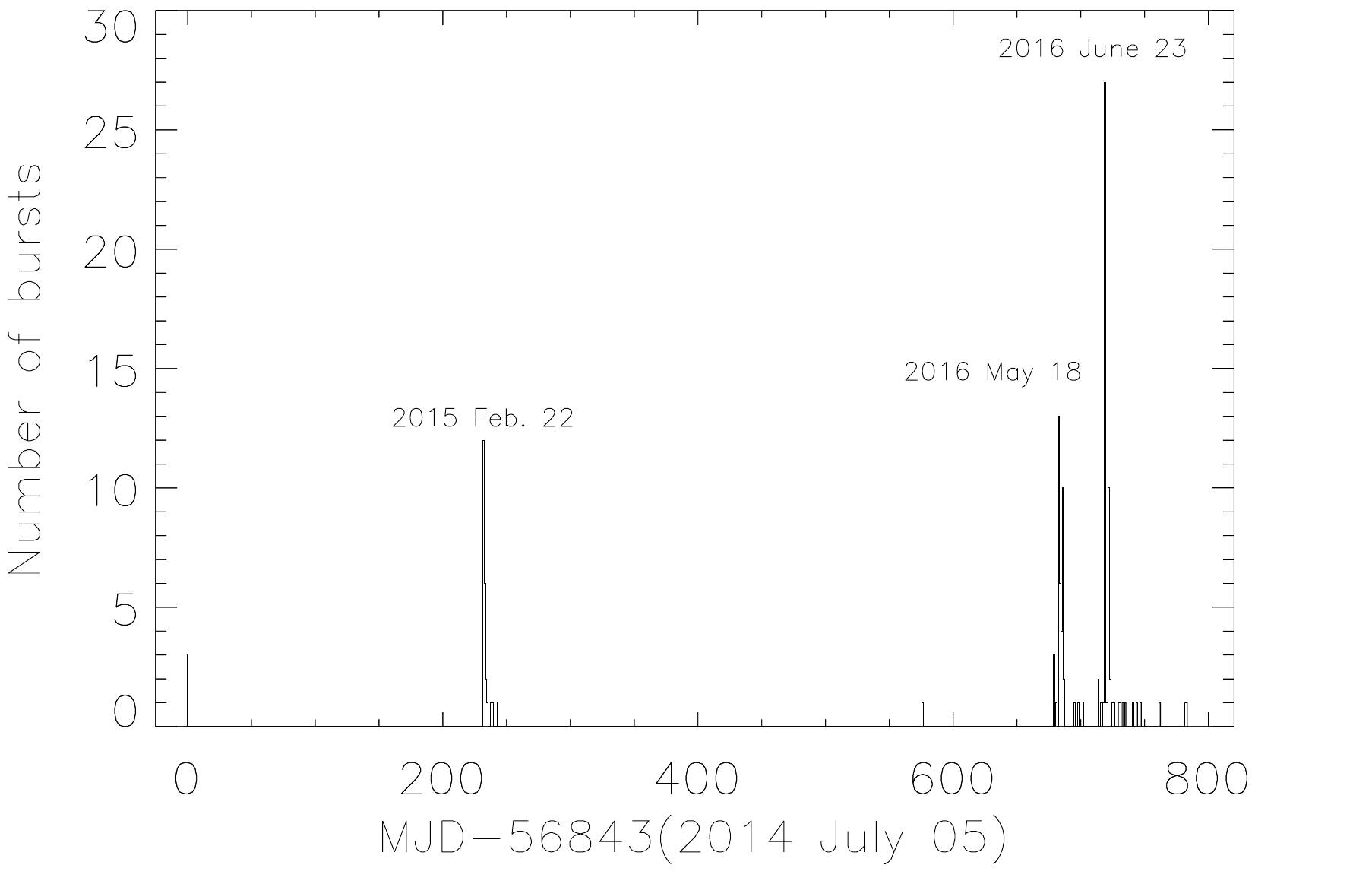}
\includegraphics*[viewport=75 160 600 495, scale=0.5]{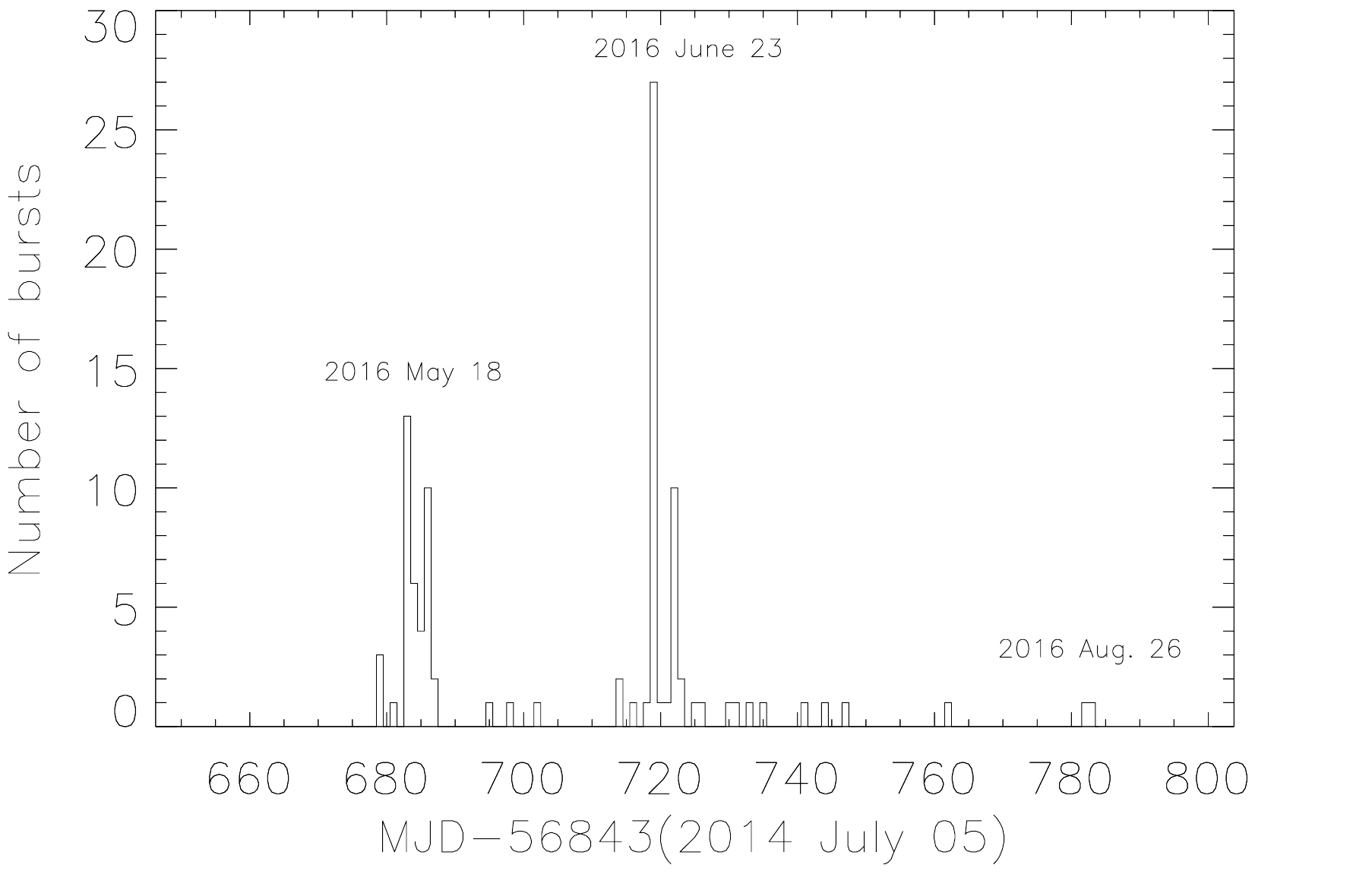}
\caption{The burst time history of \sgr in 1-day time bins. \textit{Left:} The burst history from July, 2014 - August, 2016. \textit{Right:} The expanded burst history starting from May, 2016.  \label{fig:bursthistory}}
\end{figure*}

We define an active bursting episode as the time period during which more than two bursts are emitted, with no bursts observed 10 days either side of this range. Using this definition, we find four bursting episodes for \sgrnos. We exhibit the source burst history in Figure \ref{fig:bursthistory} and summarize the four episode properties in Table \ref{tab:burstepisode}. \sgr became increasingly active in 2015 and 2016. It ceased activity after August, 2016. We notice that at least ten bursts were detected within one day for all active burst episodes, except for 2014. Considering the persistent flux increase following each episode \citep{younes2017}, \sgr should be classified as a prolific transient in the scheme of \citet{gogus2014}. 

\begin{deluxetable*}{ccccccc}
\tablenum{2}
\tablecaption{\sgr Activation Intervals. \label{tab:burstepisode}}
\tablewidth{0pt}
\tablehead{
\colhead{Episode } & \colhead{Start date} & \colhead{End date} & \colhead{in BAT/GBM/Both} & \colhead{Total Number} & \colhead{Burst fluence$^{\dagger}$} & \colhead{Burst energy$^{*,\dagger}$} \\
\colhead{} & \colhead{} & \colhead{} & \colhead{} & \colhead{} & \colhead{($10^{-7}~\rm{erg}~\rm{cm^{-2}}$)} & \colhead{($10^{39}~\rm{erg}$)} 
}
\startdata
1 & 2014-07-05 & 2014-07-05 & 1 / 0 / 2 & 3 & 1.1 & 1.1 \\
2 & 2015-02-22 & 2015-03-05 & 2 / 22 / 0 & 24 & 41.4 & 40.1 \\
3 & 2016-05-14 & 2016-06-06 & 6 / 33 / 3 & 42 & 119.5 & 115.8 \\
4 & 2016-06-18 & 2016-07-21 & 5 / 48 / 1 & 54 & 456.1 & 442.0 \\
\enddata
\tablecomments{$^*$ Assuming a distance of 9 kpc to \sgrnos. \\
$^{\dagger}$ Values are the sum of both the burst fluence and the burst energy for all bursts in each episode.} 

\end{deluxetable*}

Aside from these bursting episodes, we found four isolated events with burst IDs of 28, 125, 126 and 127)\footnote{A fourth BAT burst reported later \citep{cummings2014} is not included in the sample, as the count rate data were not sufficient for further detailed analysis.}. We also searched seven days either side of the intermediate flare that occurred on the 12$^{th}$ of April 2015, but found no other bursts.

\subsection{Burst localization}
As an imaging instrument, the BAT has the capability to locate each triggered burst to within an uncertainty of several arcminutes. We search for bursts in the mask-weighted lightcurves that trace back to the position of \sgrnos, considering all additional short events to also originate from the source (not statistical fluctuations). The false positive rate of a change point (a block containing two change points), was set to 5~\% during the search process, using the data and prior number of change points. The algorithm is defined by simulations of the pure noise \citep{scargle2013}. We iterate the search process until no further modification to the change points is necessary and the parameters are consistent. See \citet{scargle2013} for more details.

\begin{figure*}
\includegraphics*[viewport=80 140 600 490, scale=0.9]{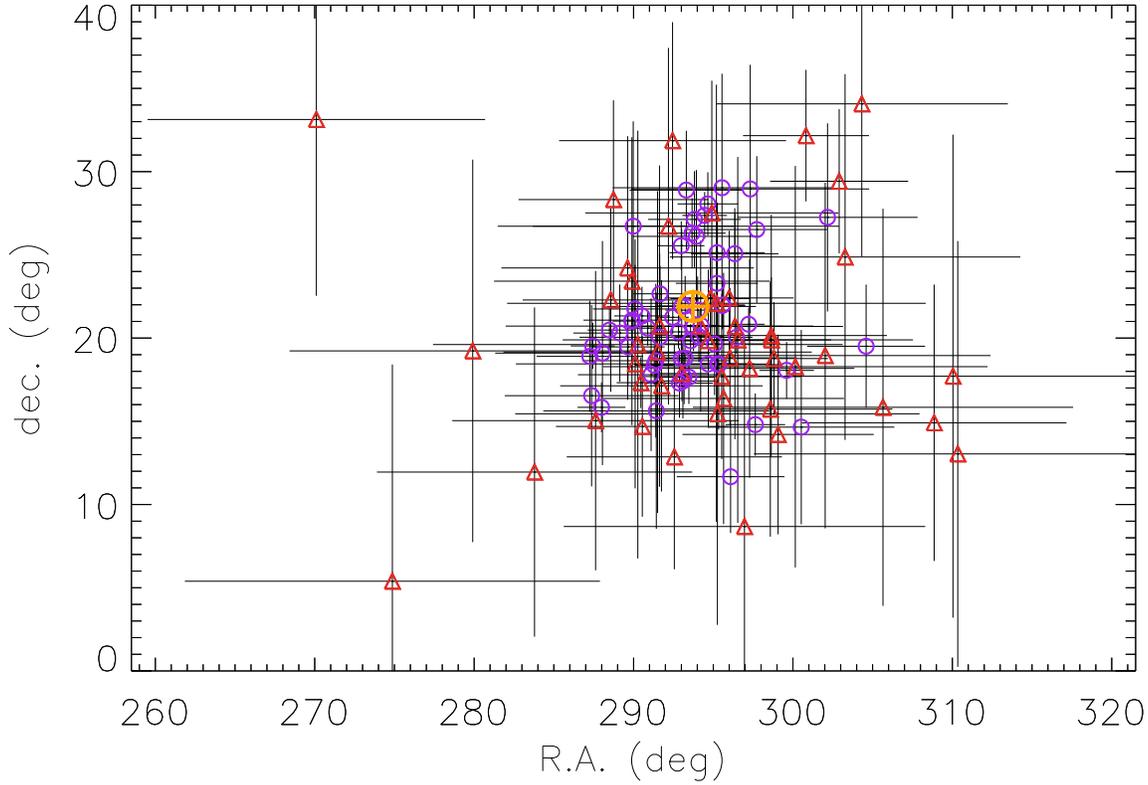}
\caption{Locations of \sgr bursts detected with GBM. The magnetar position is marked with an orange circle with a cross. Open circles and triangles represent triggered and un-triggered bursts, respectively. \label{fig:bstloc}}
\end{figure*}

\begin{figure*}
\includegraphics*[viewport=75 160 600 495, scale=0.5]{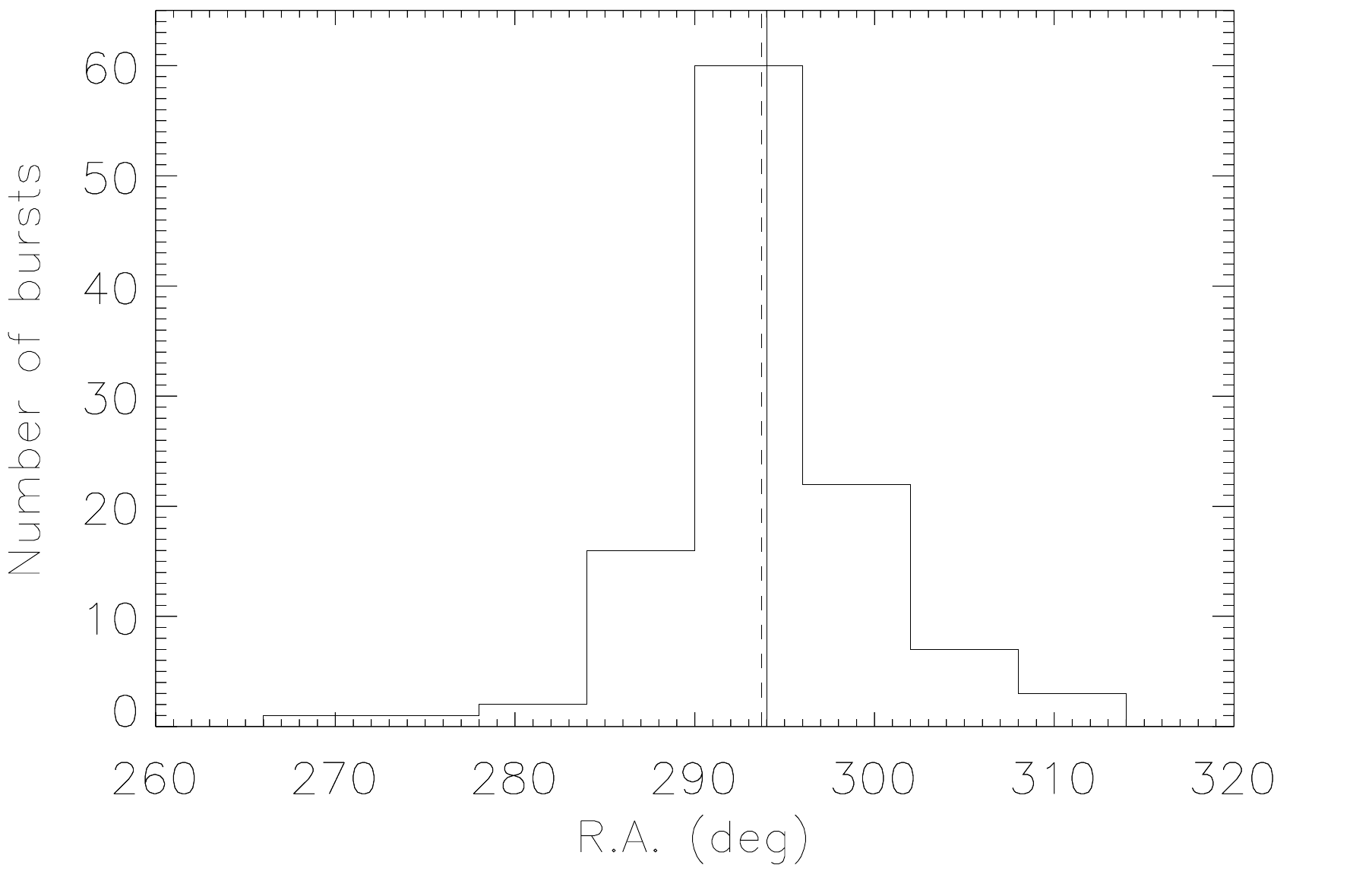}
\includegraphics*[viewport=75 160 600 495, scale=0.5]{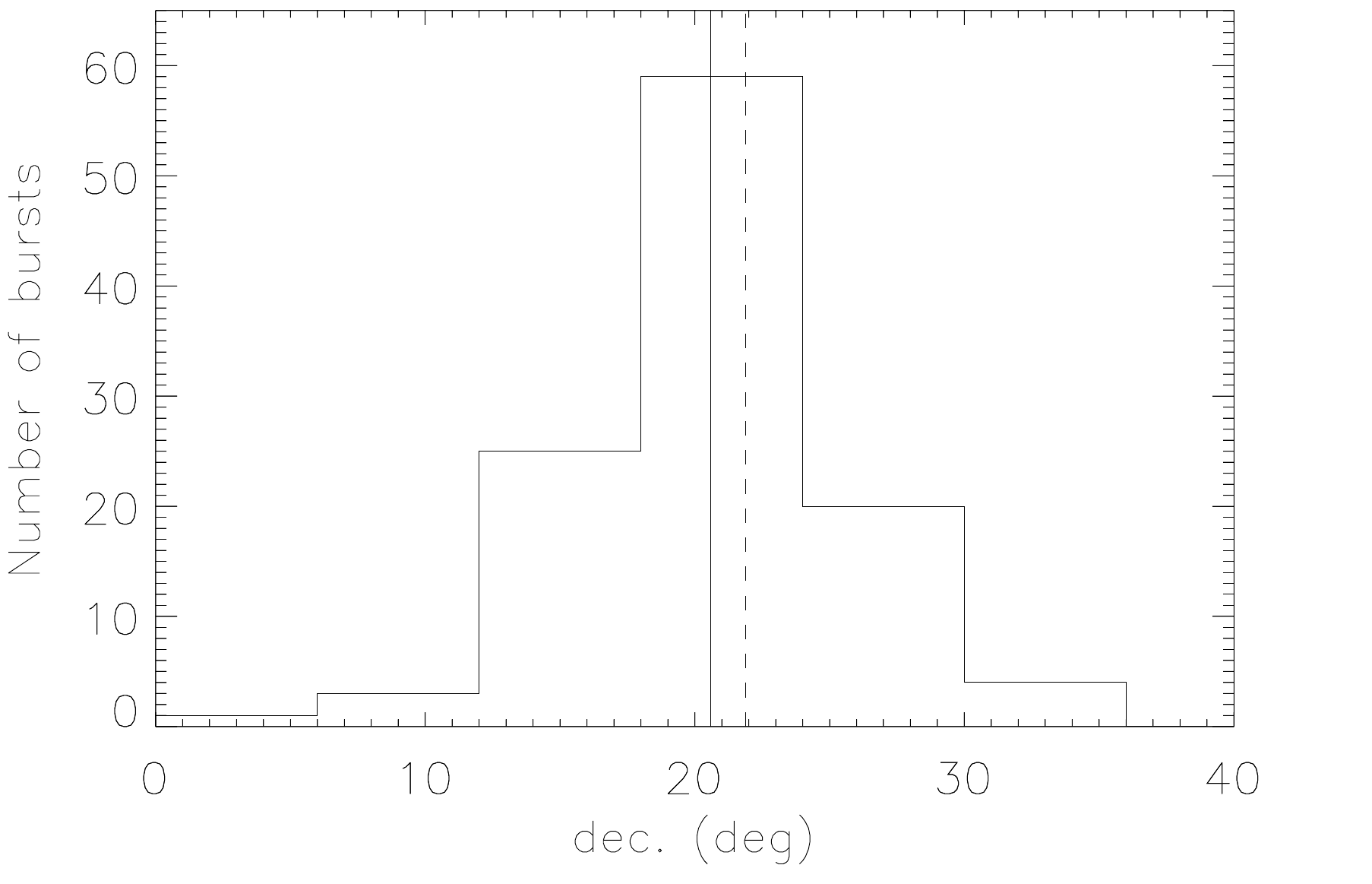}
\caption{The R.A. (\textit{left}) and Dec. (\textit{right}) distributions for the GBM burst positions. The mean values are marked as solid lines, while the dashed lines indicate the real position of \sgr in each panel. \label{fig:bstradechist}}
\end{figure*}

\textit{Fermi}/GBM provides rough burst locations by combining the count rates in the NaI(Tl) detectors that meet the aforementioned source-angle criterion of $\leq$60$^{\circ}$. The uncertainty of these locations is typically several degrees, depending on the burst peak intensity. Both triggered and un-triggered GBM events are localized with the Daughter Of Locburst (DOL) code \citep{vonkienlin2012} using counts below $50~\rm{keV}$. Table \ref{tab:burstlist} lists the measured locations and statistical errors at the $1\sigma$ confidence level\footnote{All statistical errors of the fit parameters presented in this paper are at the $1\sigma$ confidence level. }. We find that the error bars presented in Table \ref{tab:burstlist} and Figure \ref{fig:bstloc} do not include the systematic uncertainties which are at least $3^{\circ}$\citep{connaughton2015}. GBM Bursts located around the known position of \sgr (some with quite large uncertainties), are shown in Figure \ref{fig:bstloc}. Both the right ascension and declination values of bursts follow Gaussian distributions with the mean values close to the source position (Figure \ref{fig:bstradechist}). While the locations of several of these bursts are consistent with SGR$~1900+14$ (R.A.$=286.8^{\circ}$, Dec.$=9.3^{\circ}$ \citep{frail1999}), we find no observational evidence to suggest this source was active around the burst times presented in this study. Therefore, we consider all these events to be from \sgrnos.

\subsection{Burst durations}

We select three parameters to quantify each burst duration. The Bayesian block duration ($T_{\rm{bb}}$) is a direct benefit from our search process, and is the total length of all Bayesian blocks for a burst event, without any artificial selection \citep{lin2013}. We record $T_{\rm{bb}}$ for both BAT and GBM bursts. $T_{90}$ ($T_{50}$) is defined as the time interval over which the cumulative energy fluence of the burst increases from $5\%$ ($25\%$) to $95\%$ ($75\%$) of the total fluence \citep{kouveliotou1993}. These are only calculated for the GBM bursts. The speciality of $T_{90}$ (and $T_{50}$) is that when calculating energy fluence, the response of the instrument is deconvolved. However, they may be affected by the selection \citep{lin2011}.

\begin{figure*}
\includegraphics*[viewport=75 160 600 495, scale=0.5]{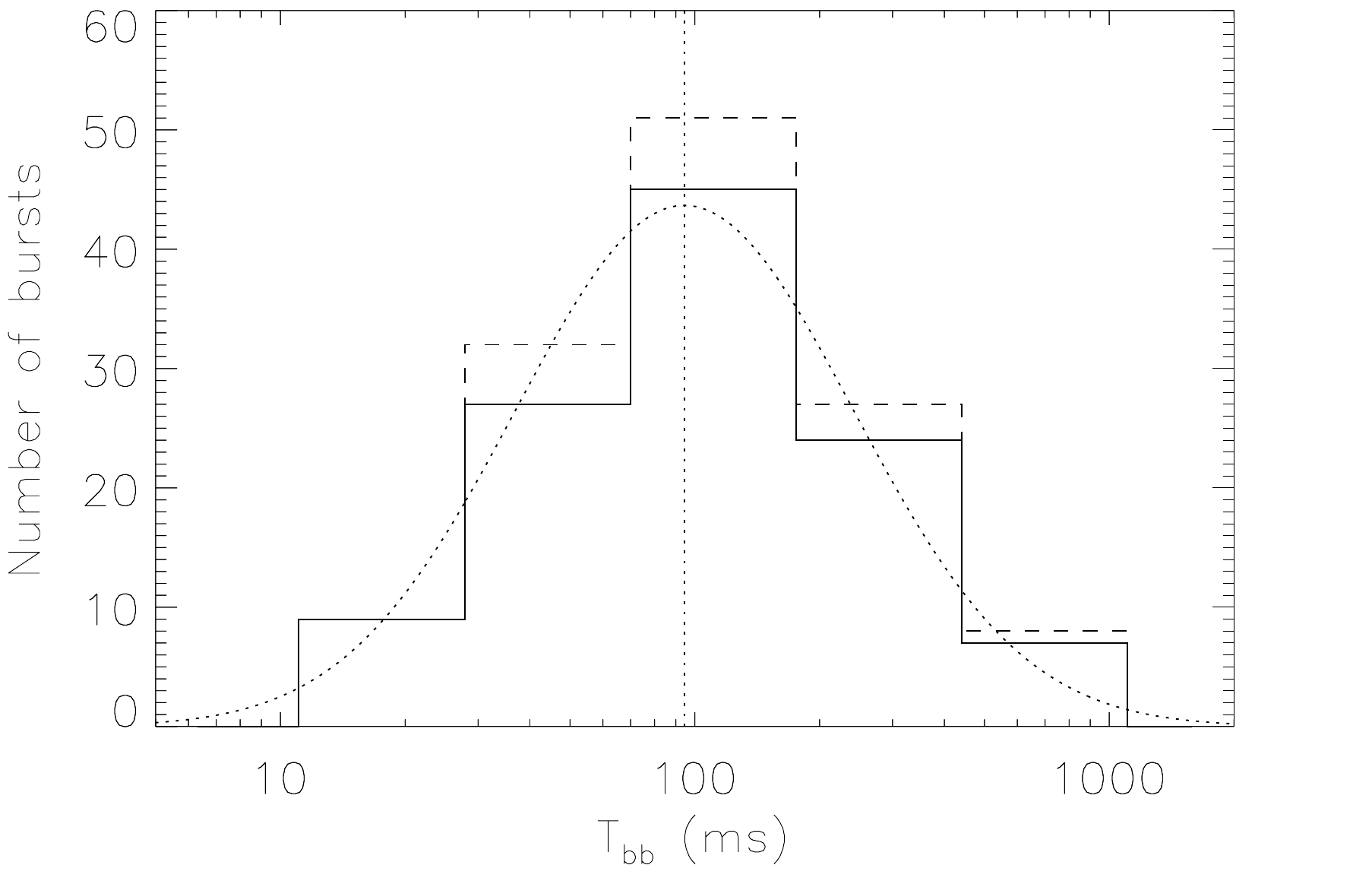}
\includegraphics*[viewport=75 160 600 495, scale=0.5]{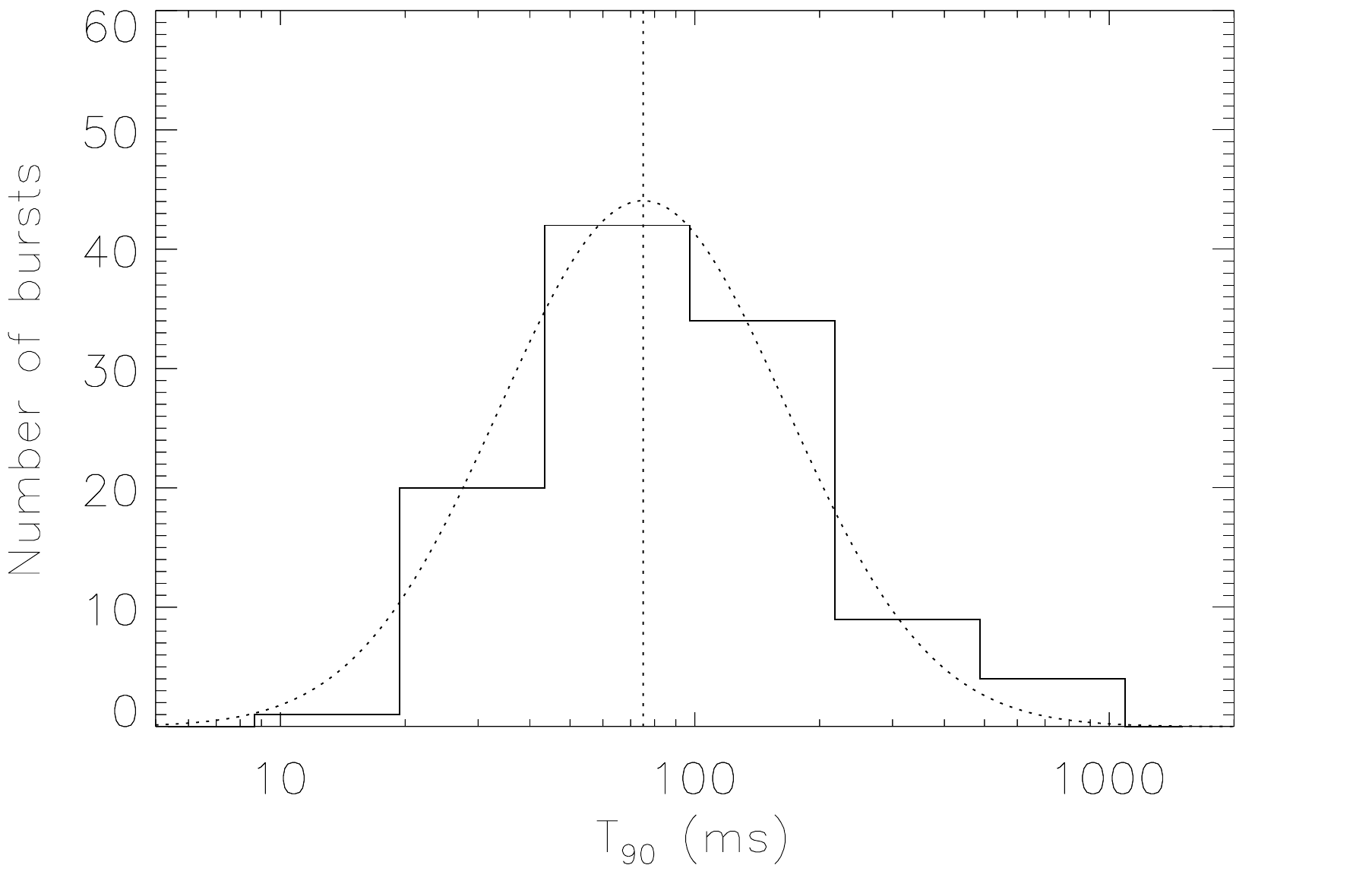}\\
\includegraphics*[viewport=75 160 600 495, scale=0.5]{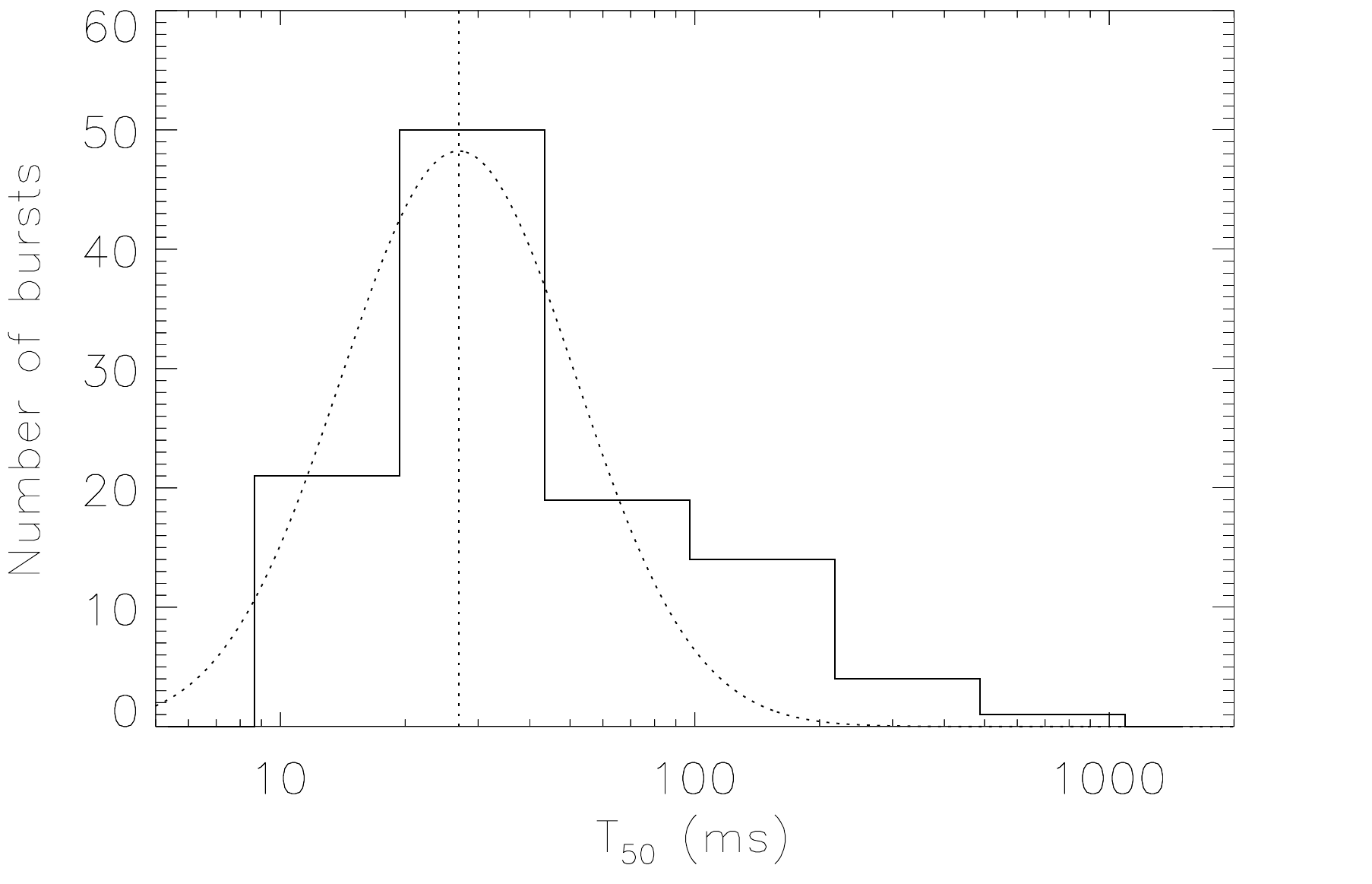}
\includegraphics*[viewport=70 140 600 490, scale=0.5]{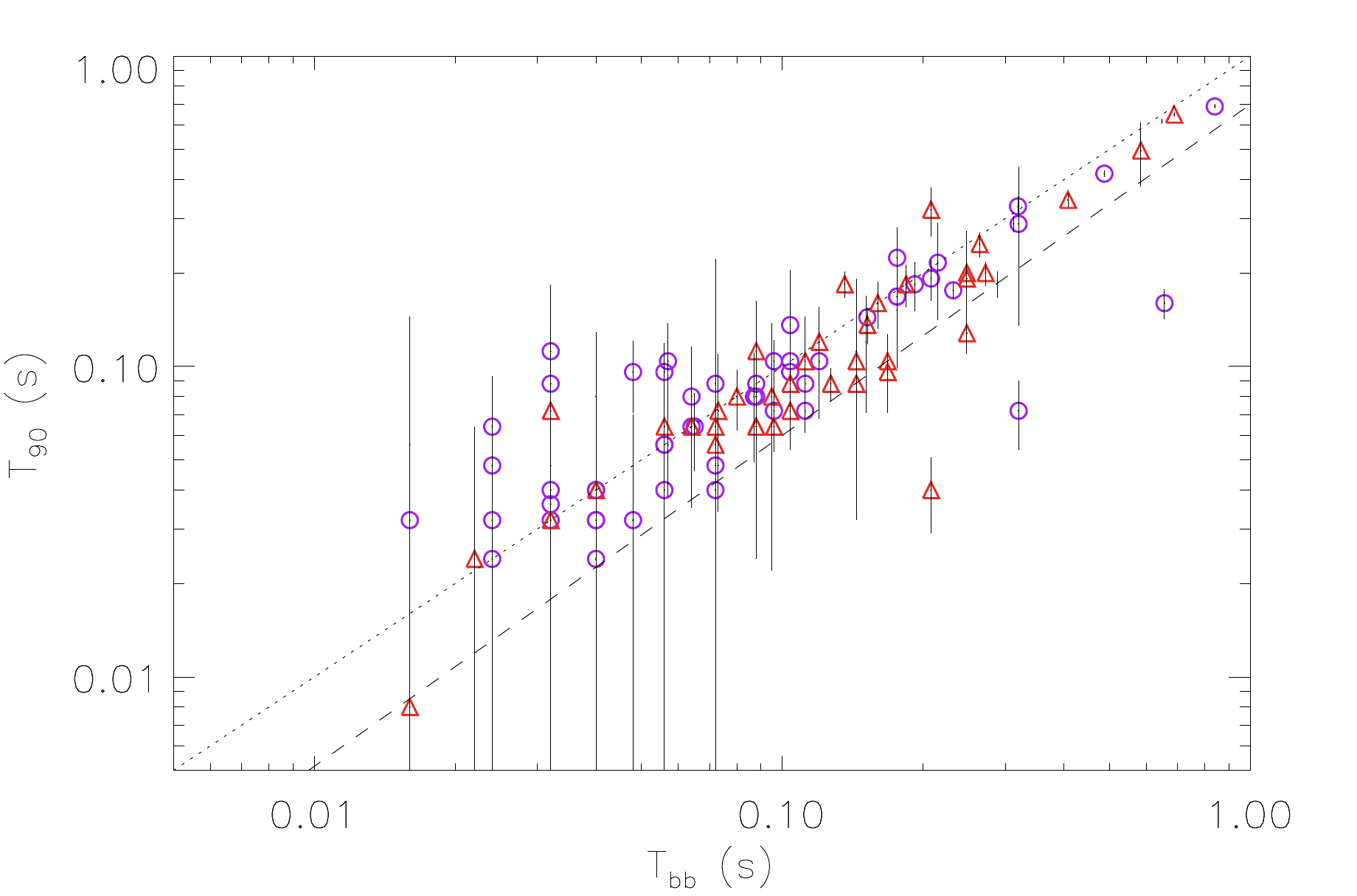}
\caption{Distributions of $T_{\rm{bb}}$ (\textit{top left}), $T_{90}$ (\textit{top right}) and $T_{50}$ ($\textit{bottom left}$) for the GBM bursts. The distribution of $T_{\rm{bb}}$ for {\it all} events is also presented in the \textit{top left} panel with a dashed histogram. The dotted curves and lines in these three panels are the best Gaussian fits to the histograms and the mean values from the fit. The correlation between $T_{90}$ and $T_{\rm{bb}}$ is shown in the \textit{bottom right} panel. The purple circles and red triangles mark GBM triggered and un-triggered events, respectively. The dashed line is the best power law fit to the correlation trend and the dotted one is the  $T_{90}=T_{\rm{bb}}$ line.  \label{fig:bstduration}}
\end{figure*}

Similar to the distribution of bursts from other magnetars \citep{collazzi2015}, $T_{\rm{bb}}$, $T_{90}$ and $T_{50}$ follow Gaussian distributions when using a logarithmic scale. We present their distributions in Figure \ref{fig:bstduration}. The best Gaussian fit parameters and their statistics to these burst durations are presented in Table \ref{tab:bstdurfit}. For GBM bursts, $T_{\rm{bb}}$ is generally longer than $T_{90}$, and the two quantities are well correlated (Figure \ref{fig:bstduration}). The corresponding Spearman's rank correlation coefficient is $0.85$, with a chance probability of $5.13\times10^{-31}$. A power law fit to the trend results in  $T_{90}\propto T_{\rm{bb}}^{0.81\pm0.01}$. A detailed list of the temporal characteristics for each burst is presented in Table \ref{tab:burstlist}.

\begin{deluxetable*}{ccccc}
\tablenum{3}
\tablecaption{\sgr Burst Duration Statistics. \label{tab:bstdurfit}}
\tablewidth{0pt}
\tablehead{
\colhead{Duration} & \colhead{Sample} & \colhead{Average} & \multicolumn{2}{l}{Gaussian fit} \\
\colhead{ } & \colhead{ } &\colhead{ (ms)} & \colhead{$\mu$ (ms)} & \colhead{$\sigma^{*}$} 
}
\startdata
$T_{\rm{bb}}$ & BAT+GBM & 138 & $93^{+4}_{-3}$  & $0.40\pm0.02$  \\
$T_{\rm{bb}}$ & GBM & 139 & $94\pm5$  & $0.41\pm0.02$ \\
$T_{90}$ & GBM & 123 & $75\pm3$ & $0.35\pm0.02$ \\
$T_{50}$ & GBM &58 & $27^{+4}_{-3}$ & $0.28\pm0.05$\\
\enddata
\tablecomments{$^*$ using a logarithmic scale }
\end{deluxetable*}

\subsection{Burst spectra}

We fit the burst spectra with five models: a power law (PL), a single blackbody (BB), an optically thin thermal bremsstrahlung (OTTB), the sum of two blackbodies (BB+BB), and a PL with an exponential cutoff at higher energies (COMPT). Earlier studies found that the two complex models (BB+BB and COMPT) are preferred in describing the spectra over broad energy ranges \citep{israel2008,lin2011,lin2012, vdh2012}. The remaining three models have been traditionally used when the data cannot constrain the parameters of more complicated models. We consider a model inadequate when at least one of its parameters enters a non-physical region above a confidence level of $1\sigma$ (e.g., a negative $kT$ or a negative $E_{\rm{peak}}$ or normalization). Of the 127 spectra, 83 can be fit well with at least one complex model. More specifically, 36 can only be fit with a BB+BB model, and three only with COMPT. Forty-four can be fit with both BB+BB and COMPT, of which 38 resulted in a smaller \textit{c-stat} (for GBM spectra) or $\chi^2$ (for BAT spectra) when fit with a BB+BB model, and the remaining 6 when fit with a COMPT model. We note that BB+BB and COMPT are not nested models and therefore, we cannot easily compare their fit statistics. Moreover, simulation results from earlier works \citep{lin2011,vdh2012} indicate that over an energy range of $8-200~\rm{keV}$, neither the BB+BB nor the COMPT model are preferred in terms of goodness of fit. The remaining 44 bursts, for which the complex models are inadequate, can only be fit with simpler models, such as a PL, a BB, or an OTTB model. We present the spectral parameters for all events in Table \ref{tab:burstspec}. In the last column of Table \ref{tab:burstlist}, we also include each burst fluence over $8-200~\rm{keV}$ for GBM spectra, and $15-150~\rm{keV}$ for BAT spectra, calculated using the spectral model with the best statistics. 

\begin{figure*}
\includegraphics*[viewport=75 170 600 495, scale=0.5]{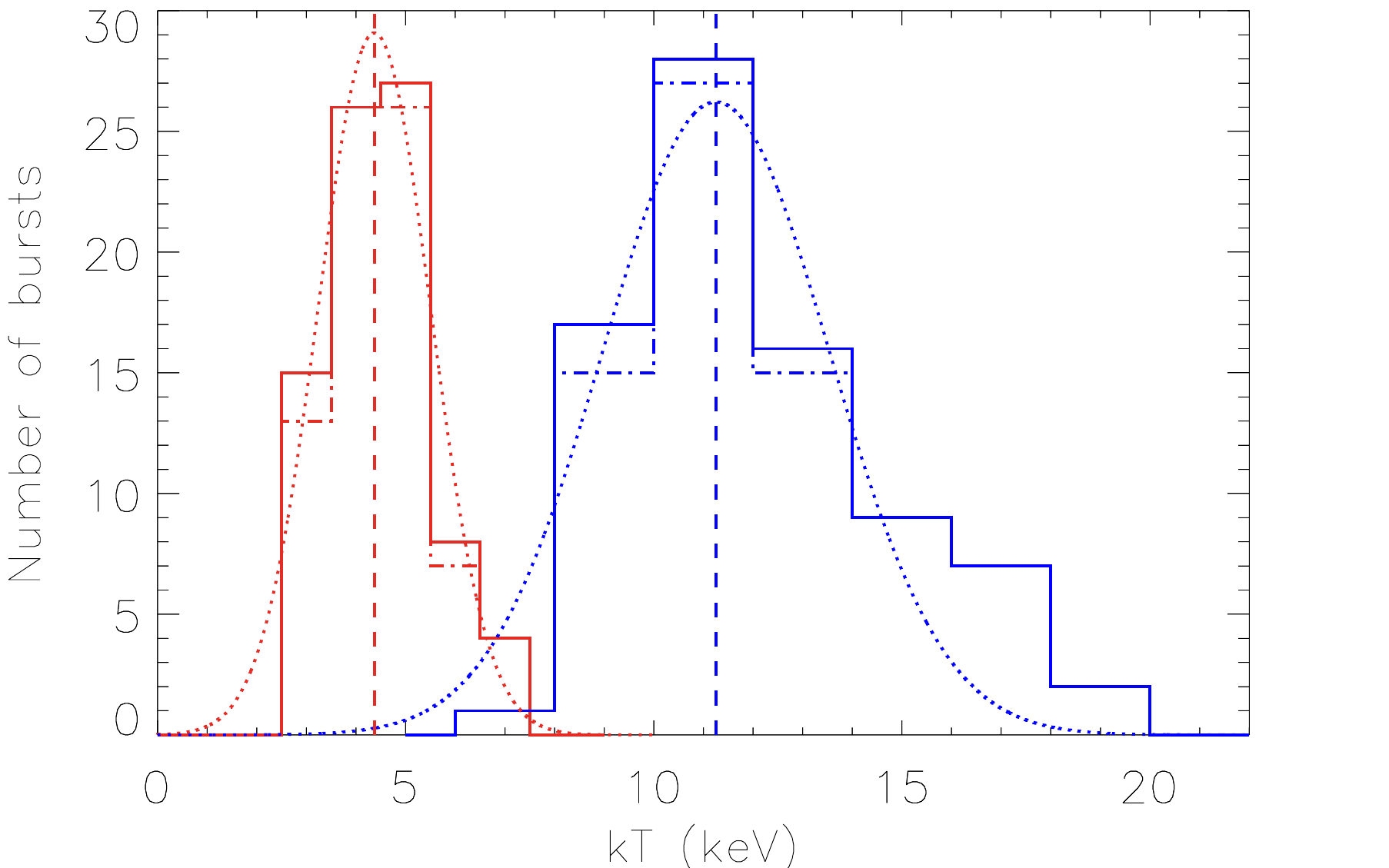}
\includegraphics*[viewport=75 170 600 495, scale=0.5]{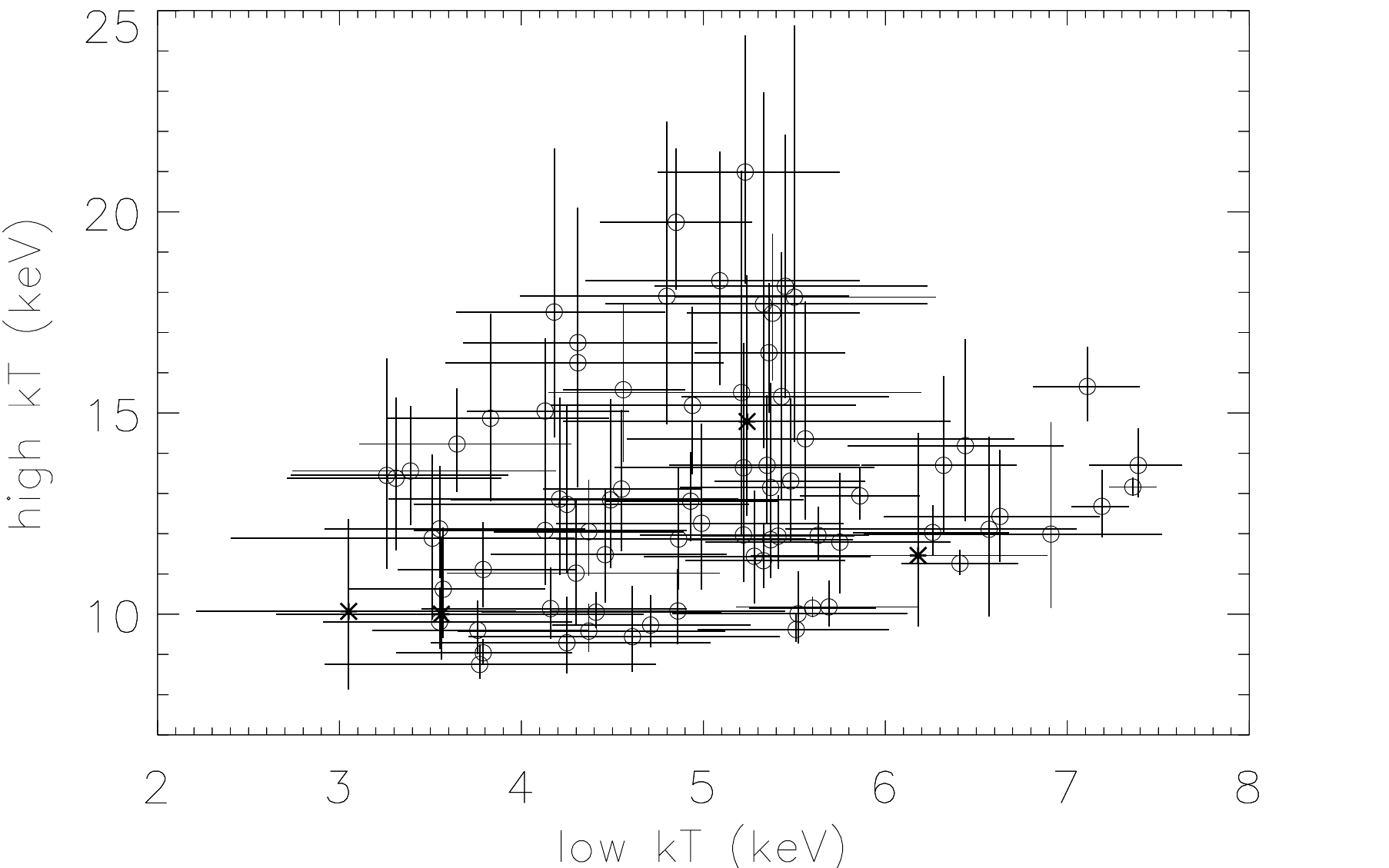}\\
\caption{\textit{Left:} The distribution for the two BB temperatures from the BB+BB fit. Solid and dashed-dot histograms are bursts detected with \textit{Swift}/BAT and GBM, respectively; dotted curves are Gaussian fits to the histograms; dashed lines mark the mean fit values. The set of red lines represents the lower BB temperature ($kT_1$), while the blue represents the higher BB temperature ($kT_2$). \textit{Right:} The high $v.s.$ the low BB temperature. Open circles represent GBM bursts and crosses are bursts detected only with the BAT.   \label{fig:specBBBB}}
\end{figure*}

\begin{figure*}
\includegraphics*[viewport=50 140 600 490, scale=0.45]{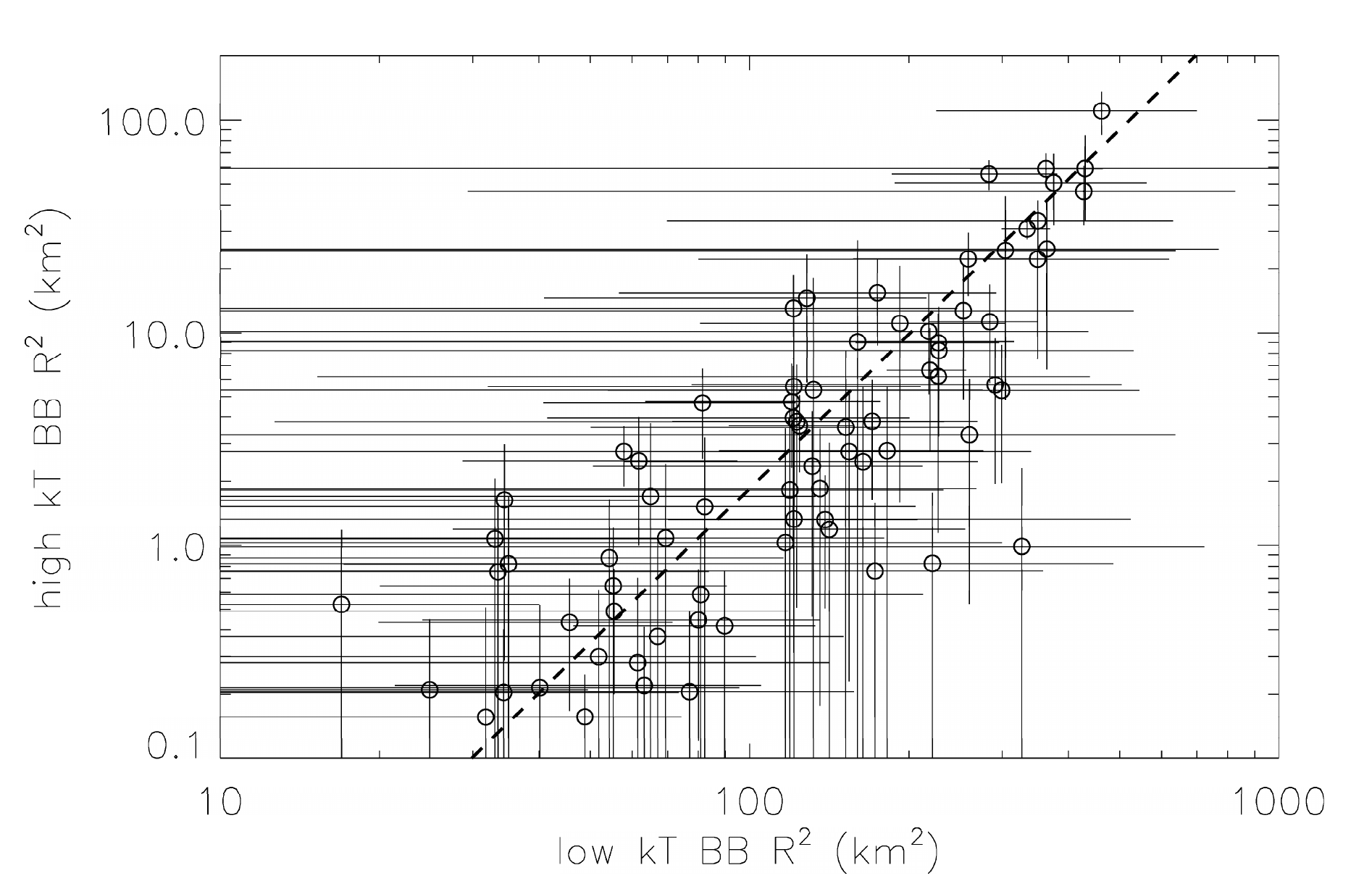}
\includegraphics*[viewport=50 140 600 490, scale=0.45]{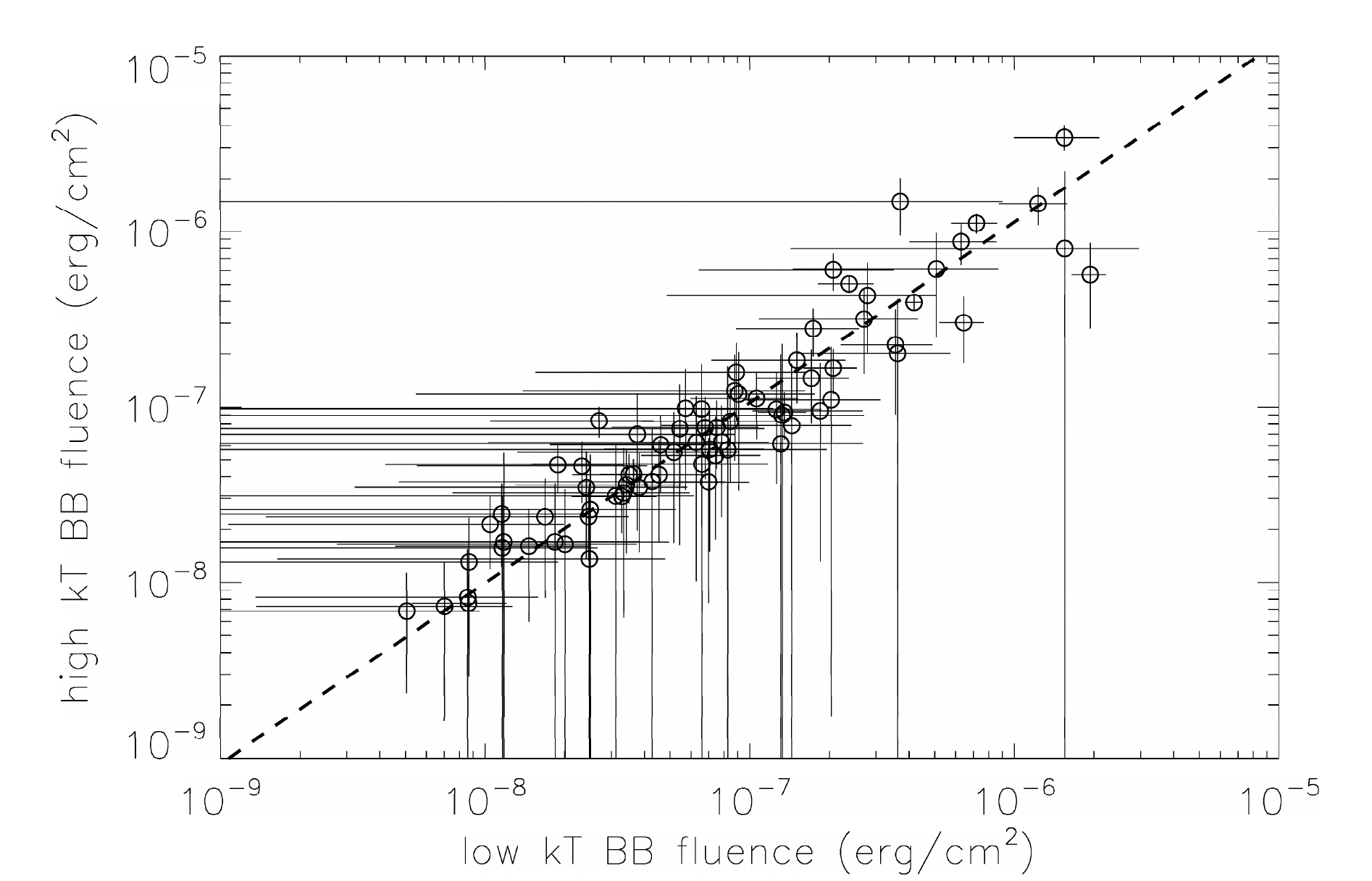}\\
\includegraphics*[viewport=50 140 600 490, scale=0.45]{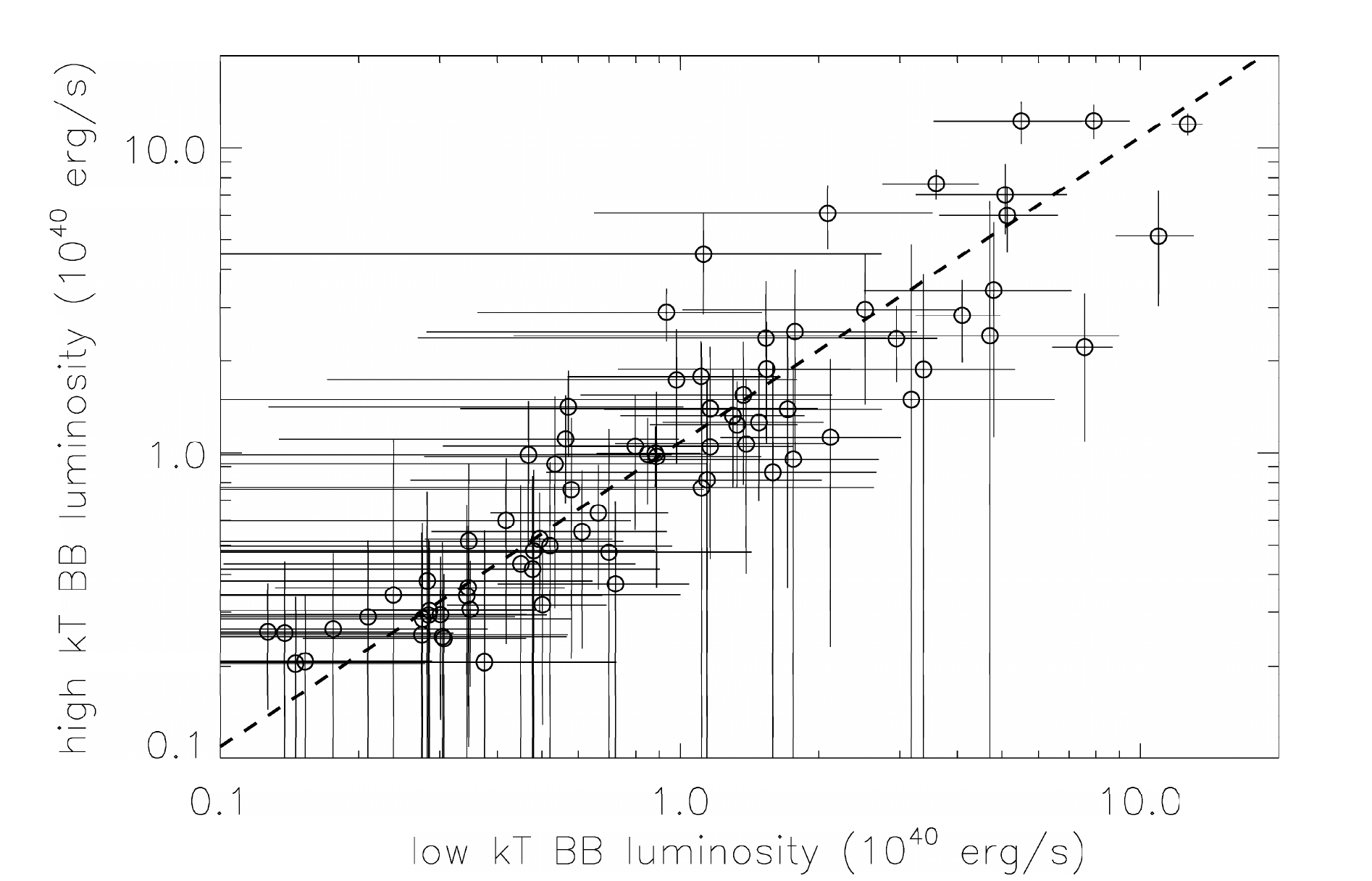}
\includegraphics*[viewport=50 140 600 490, scale=0.45]{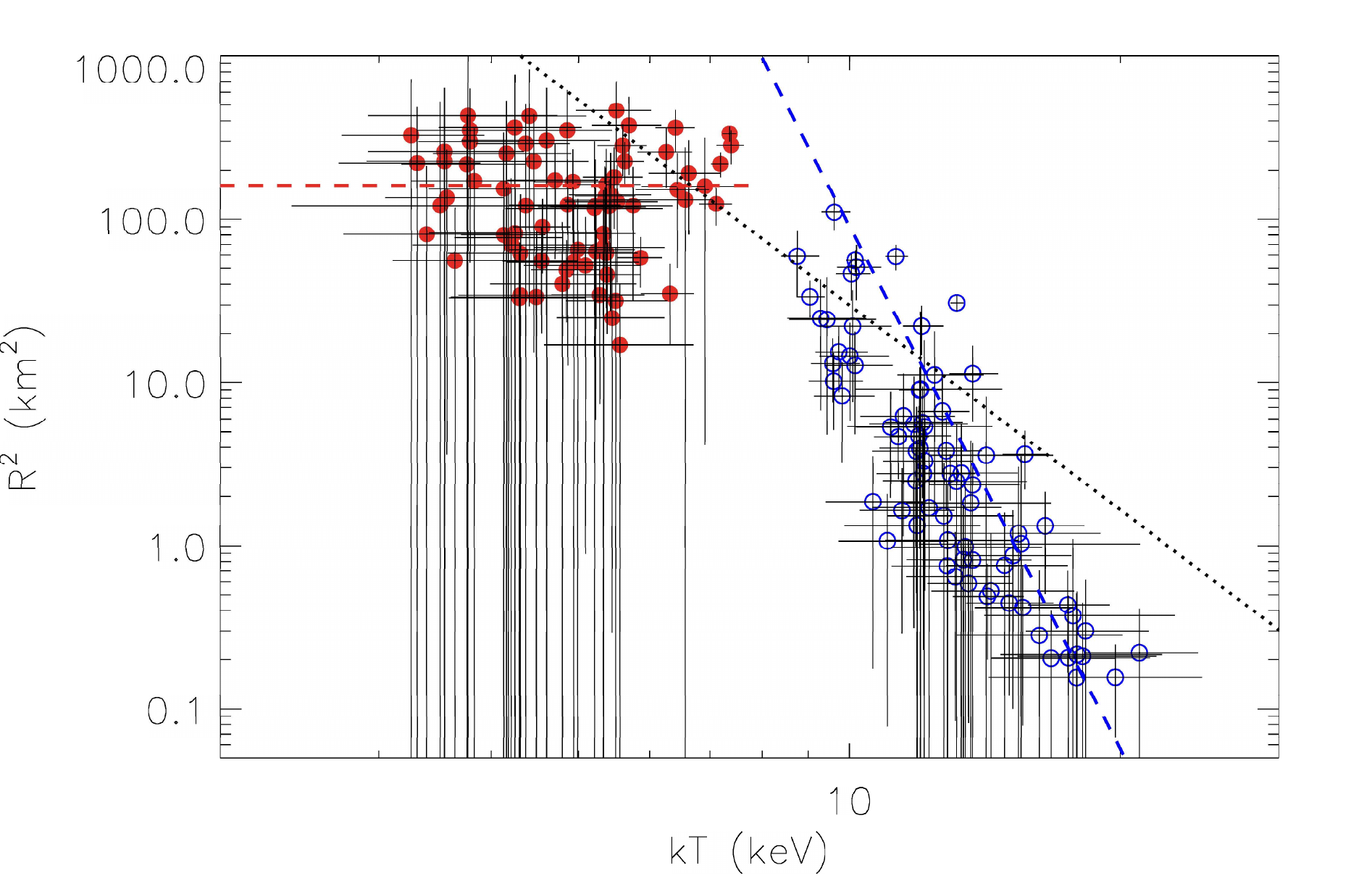}
\caption{Correlations of the emission areas (\textit{top left}), fluences (\textit{top right}) and luminosities (\textit{bottom left}) of the two BBs. The dashed lines are the best fit PLs for each correlation. \textit{Bottom right:} the BB emission areas ($R^2$) as a function of the two BB temperatures. The red dots and blue circles represent the cool and hot BB temperatures ($kT_1$ and $kT_2$), respectively. The blue dashed line is the PL that best fits to the hot BB component trend, while the red dashed line represents the mean value of the cool BB emission area. The dotted line is the PL that best fits both BB components.   \label{fig:BBBBcor}}
\end{figure*}

Using the parameters of our BB+BB fits to 80 bursts (76 fit with GBM data and 4 fit with BAT data), we find that both low and high BB temperatures follow a Gaussian distribution with a mean value of $4.4\pm0.1~\rm{keV}$ ($\sigma=1.1\pm0.1~\rm{keV}$) and $11.3\pm0.4~\rm{keV}$ ($\sigma=2.3\pm0.4$) respectively, as shown in Figure \ref{fig:specBBBB}. These values do not change significantly if the BAT bursts are excluded. The hot BB temperature range is wider than the lower termperature one. The emission areas ($R^2$), energy fluences and luminosities \footnote{Assuming a distance to \sgr of 9 kpc.} of the two BB components, are strongly correlated (Figure \ref{fig:BBBBcor}). We study these correlations using only the 76 GBM bursts, as the fluxes in GBM and BAT are over different energy ranges. The Spearman rank order correlation test yields coefficients and chance probabilities of 0.8 and $1.4\times10^{-19}$ for emission areas, 0.9 and $1.9\times10^{-30}$ for luminosities, and 0.9 and $1.2\times10^{-37}$ for fluences. Notice that the errors for these quantities are not included in the correlation test. We also fit the three correlations with a PL and find that the best fit indices are $2.41\pm0.56$ for the emission areas, $1.00\pm0.12$ for the luminosities, and $1.03\pm0.12$ for the fluences. 

\begin{figure*}
\includegraphics*[viewport=50 140 620 490, scale=0.45]{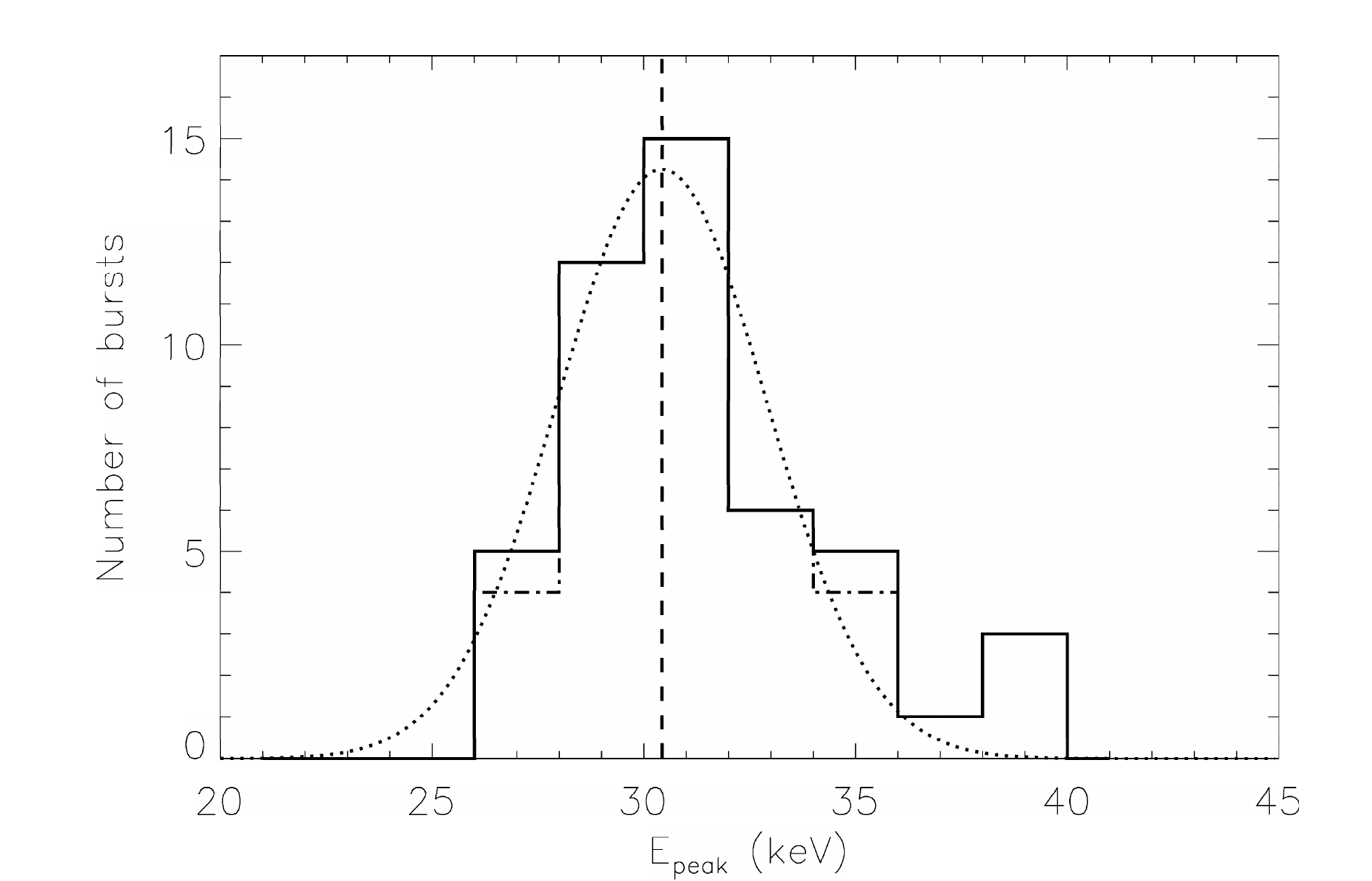}
\includegraphics*[viewport=70 140 620 490, scale=0.45]{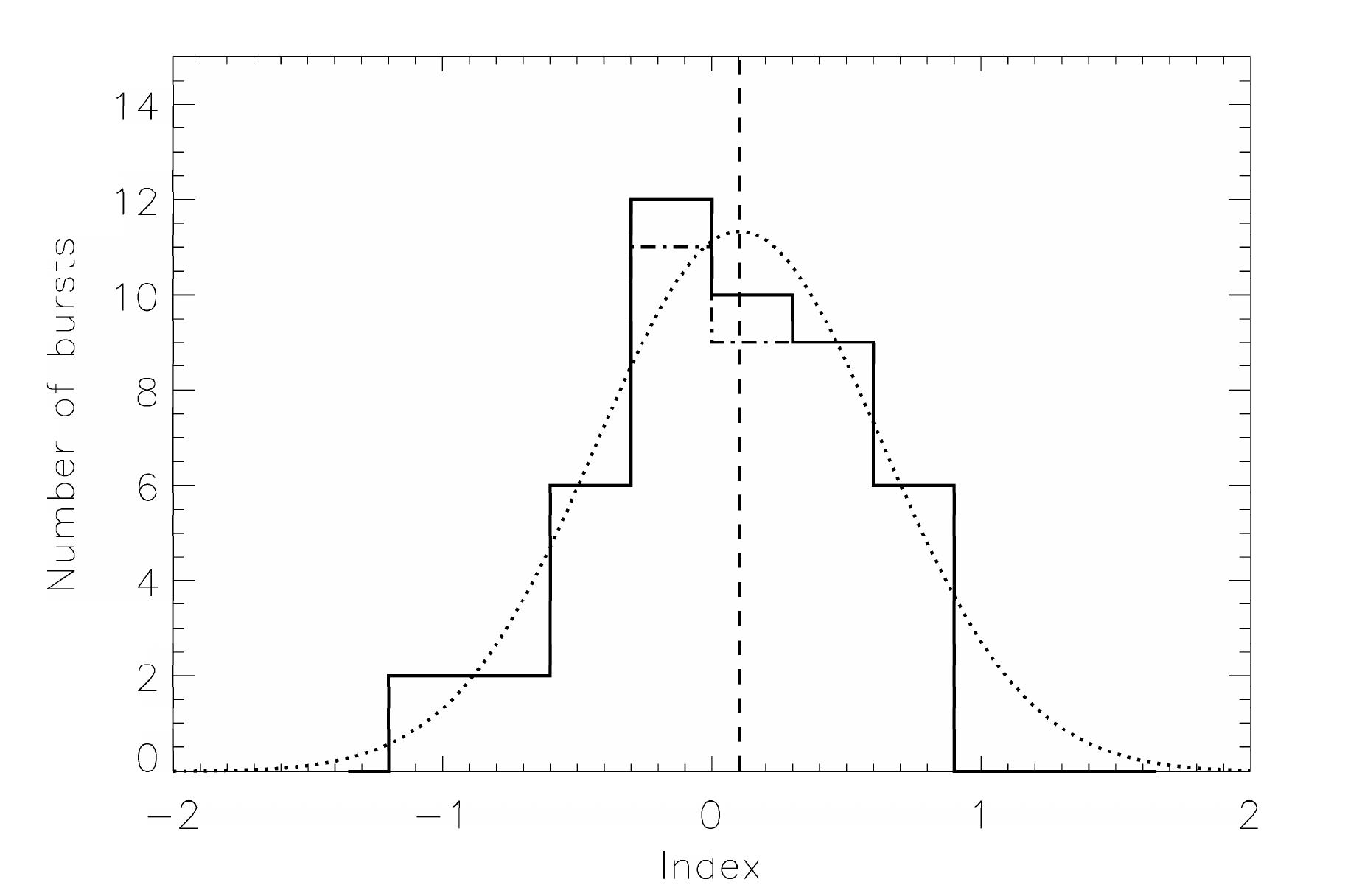}
\includegraphics*[viewport=50 140 620 490, scale=0.45]{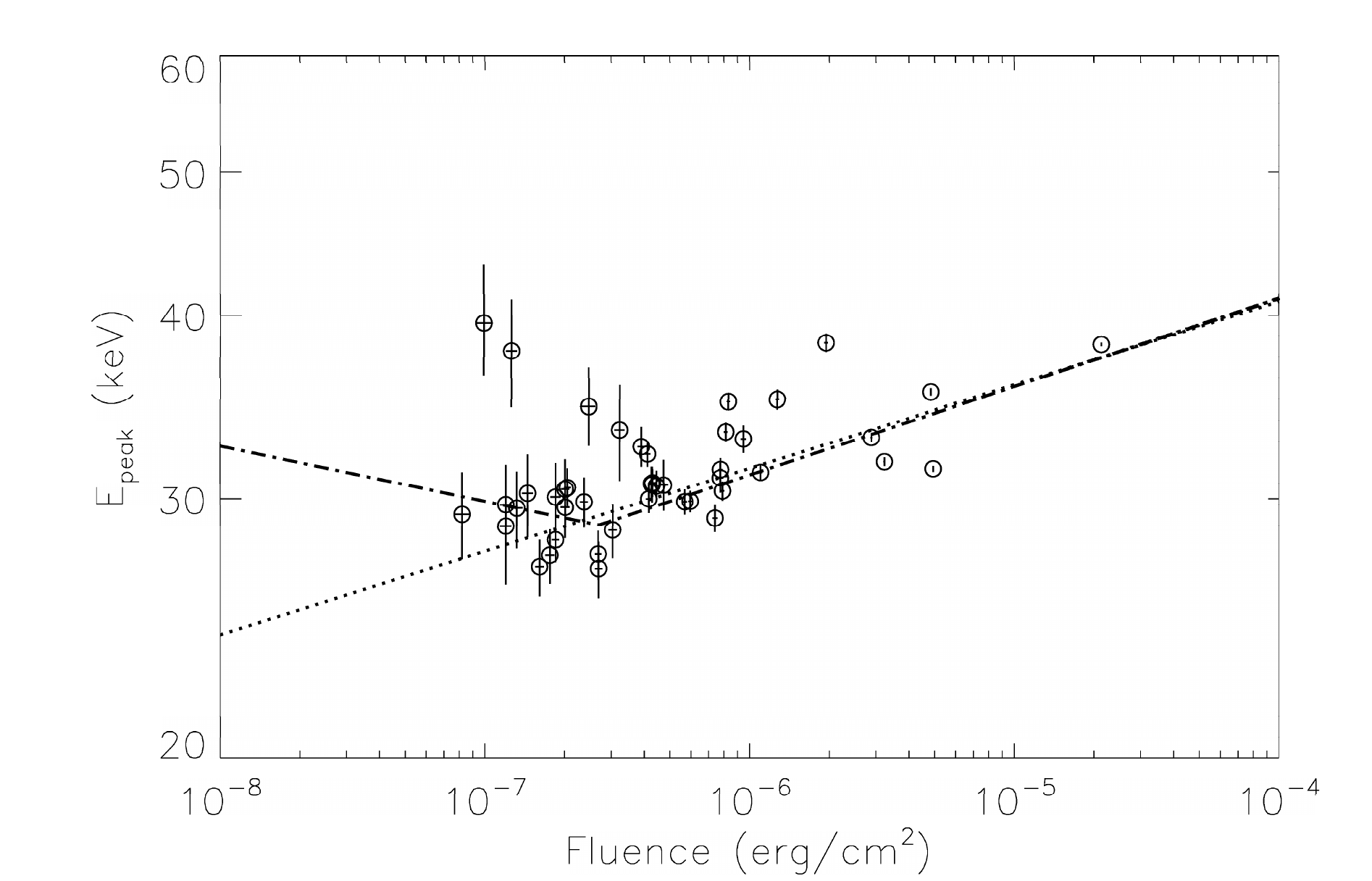}
\includegraphics*[viewport=50 140 620 490, scale=0.45]{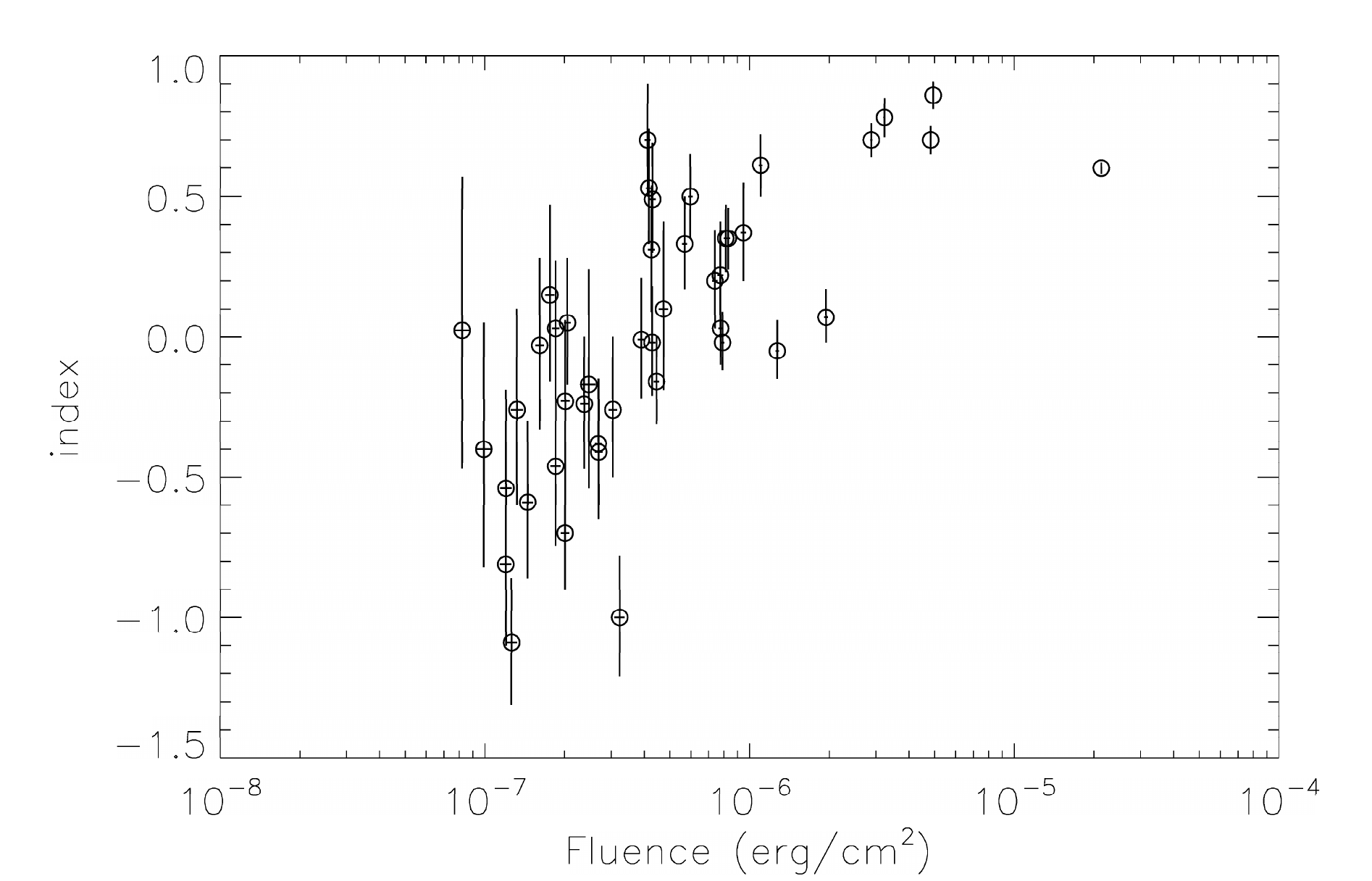}
\caption{The distributions of $E_{peak}$ (\textit{left}) and index (\textit{right}) of the COMPT model fits for 47 bursts (solid line) and 45 GBM bursts (dash-dotted line) are presented in the top left panel. The dotted curves are Gaussian fits to the histograms; dashed lines are mean values. In the lower panels, we show the evolution of $E_{peak}$ (\textit{left}) and index (\textit{right}) as a function of fluence for the 45 GBM bursts. The best PL fit to the correlation between $E_{peak}$ and fluence is shown as a dotted line, while the dash-dot line shows a broken PL fit to the data. \label{fig:specCOMPT}}
\end{figure*}

We then study the distribution of the COMPT model parameters for the 45 GBM bursts (the two BAT only events are excluded from the correlation analysis due to the different energy range of the instruments). Our results are presented in Figure \ref{fig:specCOMPT}. Peak energies ($E_{peak}$) range from $\sim25~\rm{keV}$ to $\sim40~\rm{keV}$ with a mean value of $31.4~\rm{keV}$.  A Gaussian fit to the distribution gives a mean value of $30.4\pm0.2~\rm{keV}$ and $\sigma=2.5\pm0.2~\rm{keV}$. As the burst fluence increases, $E_{peak}$ becomes slightly harder. A simple PL with an index of $0.06\pm0.003$, best fits this correlation. We note that weaker events are  further from the PL fit. Similar to other Magnetars, we fit a broken PL to the data and obtain indices of $0.06\pm0.001$ and $-0.04\pm0.04$. The intersection of the two PL fits is at a fluence of $2.7\pm0.5\times10^{-7}~\rm{erg~cm^{-2}}$. We also note that the index for the lower fluence bursts is weakly constrained as the break is quite close to the lowest fluence in our sample. 
The COMPT PL index ranges from $-1$ to 1 with an average of 0.03. We also fit the index distribution with a Gaussian shape, which gives  a mean of $-0.1\pm0.1$ with $\sigma=0.5\pm0.1$. We find that the COMPT PL index is correlated with the fluence. The Spearman test yields a correlation coefficient of 0.7 with a chance probability of $8.7\times10^{-9}$. This correlation indicates that the weaker bursts have a softer spectrum.

Finally, all bursts that can only be fit with simple models are quite dim. The highest fluence among these is $1.6\times10^{-7}~\rm{erg}~\rm{cm}^2$. The average temperature for 23 OTTB bursts and 9 BB bursts is $31.2~\rm{keV}$ and $9.5~\rm{keV}$, respectively, and the average PL index for 12 PL bursts is $-1.95$.  

\section{Discussion} \label{sec:dis}

\subsection{Burst energetics}

Since its re-emergence from quiescence, \sgr exhibited four active burst episodes from 2014 through late-2016. We analyzed in detail the temporal and time-integrated spectral properties of 127 short Magnetar bursts observed with the \textit{Swift}/BAT and the \textit{Fermi}/GBM. This sample includes events with fluences ranging from $10^{-8} - 2\times10^{-5}~\rm{erg}~\rm{cm^{-2}}$ over an energy range of $8-200~\rm{keV}$. This range is comparable to that of other magnetars observed by GBM (e.g., SGR\,J$1550-5418$: \citet{vdh2012,collazzi2015} and SGR\,J$0501+4516$: \citet{lin2011}). The total energy fluence emitted in our burst sample is $6.2\times10^{-5}~\rm{erg}~\rm{cm^{-2}}$, corresponding to $1.5\times10^{39}~\rm{erg}$ under the assumption of a source distance of 9 kpc.

\begin{figure*}
\includegraphics*[viewport=70 140 600 490, scale=0.9]{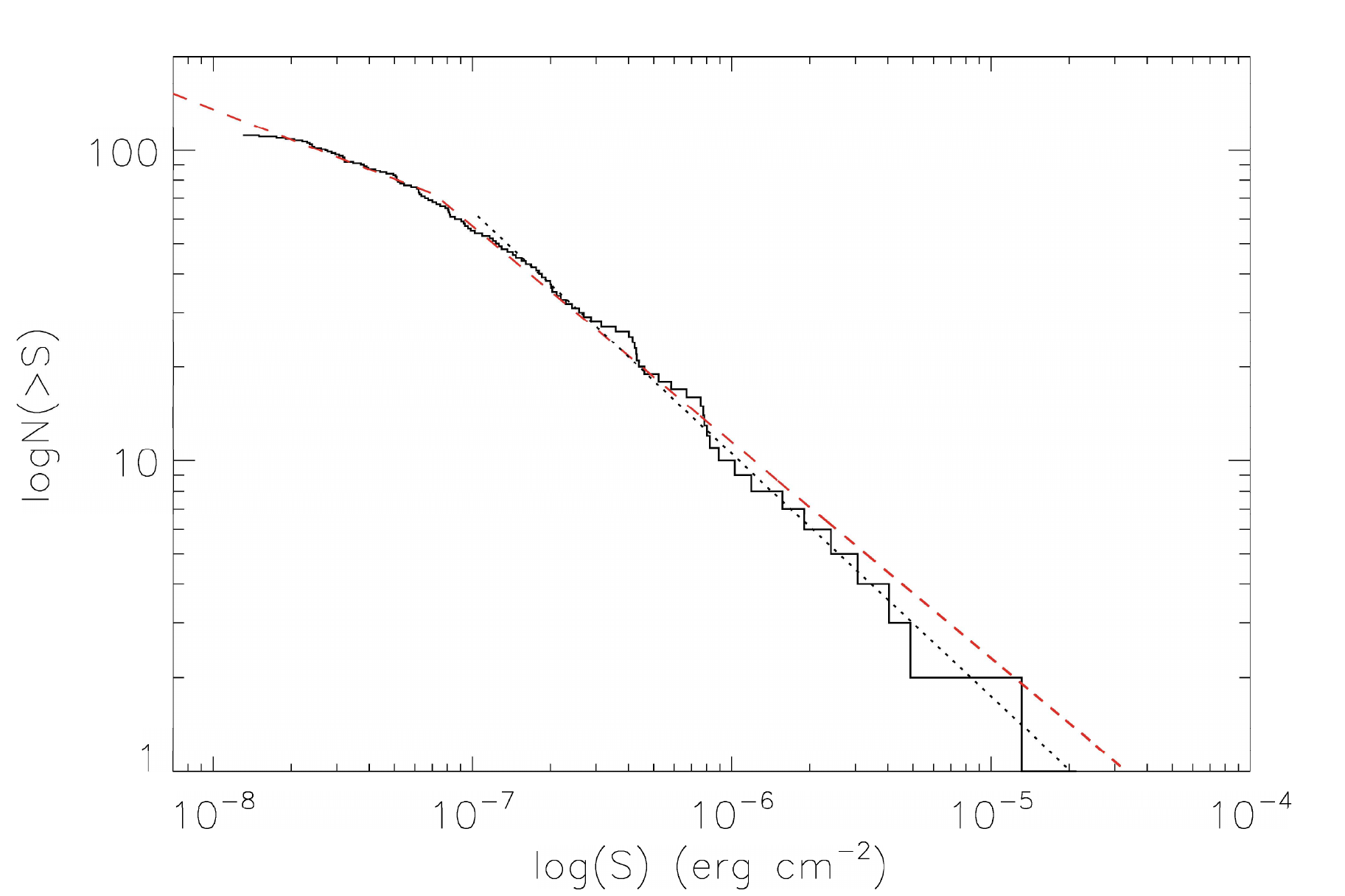}
\caption{The cumulative energy fluence distribution of \sgr bursts. The red dashed line is the best fit of the distribution with a broken PL. The dotted line is the best PL fit to the distribution above $1\times10^{-7}~\rm{erg}~\rm{cm^{-2}}$.  \label{fig:lgnlgs}}
\end{figure*}

In Figure \ref{fig:lgnlgs} we present the cumulative energy fluence ($S$) distribution for 112 GBM bursts from \sgrnos. We fit the distribution with a broken PL. The best fits to the index for the lower and higher fluences are $0.32\pm0.04$ and $0.70\pm0.03$ respectively, while the break fluence is at  $7.1\pm0.8\times10^{-8}~\rm{erg}~\rm{cm}^{-2}$. The intersection of the two PL fits is much smoother than a point, which may be due to the drop-off in the detection efficiency of the instruments and the search process. In earlier studies using the GBM data, the lower cutoff in fluence was set at $1\times10^{-7}~\rm{erg}~\rm{cm}^{-2}$ \citep{vdh2012,collazzi2015}. For comparison reasons, we select the bursts with fluences higher than $1\times10^{-7}~\rm{erg}~\rm{cm}^{-2}$, and fit their distribution with a PL model. The index that best fits the data is $N(>S) \propto S^{-0.78\pm0.01}$, which is comparable to the value reported for other Magnetars \citep{cheng1996,collazzi2015}. The differential distribution of burst fluences, $dN/dE \propto E^{-1.78}$,  is consistent with the `Gutenberg-Richter' PL for earthquakes in different active regions, as pointed out by \citet{cheng1996}. The similarity between these short magnetar bursts and earthquakes supports the hypothesis that short bursts from magnetars are due to the sudden release of energy from cracks in the solid crust of neutron stars \citep{dt1992}. We also need to keep in mind that the PL distribution is characteristic of self-organized criticality systems \citep{esposito2018} and exclusive properties are therefore crucial to prove the origin of short bursts.

\subsection{Bursts and outbursts}

About $97\%$ of our burst sample happened during four active episodes. From the first to the forth episode, the number of bursts increased by an order of magnitude, while the energy released by these bursts increased by three orders of magnitude (Table \ref{tab:burstepisode}). Following the onset of each episode, several X-ray instruments, such as XRT, \textit{XMM-Newton}, \textit{Chandra} and \textit{NuSTAR}, observed the persistent emission outburst of \sgrnos. \citet{younes2017} performed detailed analyses from these observations and concluded that the energy released in the persistent outburst of \sgr is roughly the same order of magnitude for all four episodes. The energy ratio between the burst and persistent emission in outbursts is $\sim 0.03$, $\sim 0.66$, $\sim 5.79$ and $\sim 12.28$ for active episodes one to four, respectively. Note that this ratio is a lower limit due to the incompleteness of the burst detection. 

As a transient Magnetar, the persistent flux increase of \sgr is modest at the onset of each outburst \citep{coti2018}. Its value changed by factors of 5$-$10, while most transient Magnetars exhibited X-ray flux increases of $\sim50-100$ at the activation onset. This increase was also usually coincident with one or more bursts. The rapid increase in X-ray flux is attributed to the cooling of a heated crustal zone at the start of the outburst \citep{lyubarsky2002}.

We can infer from its flux increase that crustal heating takes place in \sgr. However, as magnetar bursts are likely to radiate energy efficiently, we assume only a small fraction of energy is left to heat the crust. This could explain why the outbursts of \sgr are not bright at the onset, or over a longer interval. We note that this could also be linked to its recurring outburst behavior, as the source flux drops to near the quiescent flux level quickly over several months and therefore every burst reactivation would initiate a new outburst episode.

\begin{figure*}
\includegraphics*[viewport=75 165 600 495, scale=1.0]{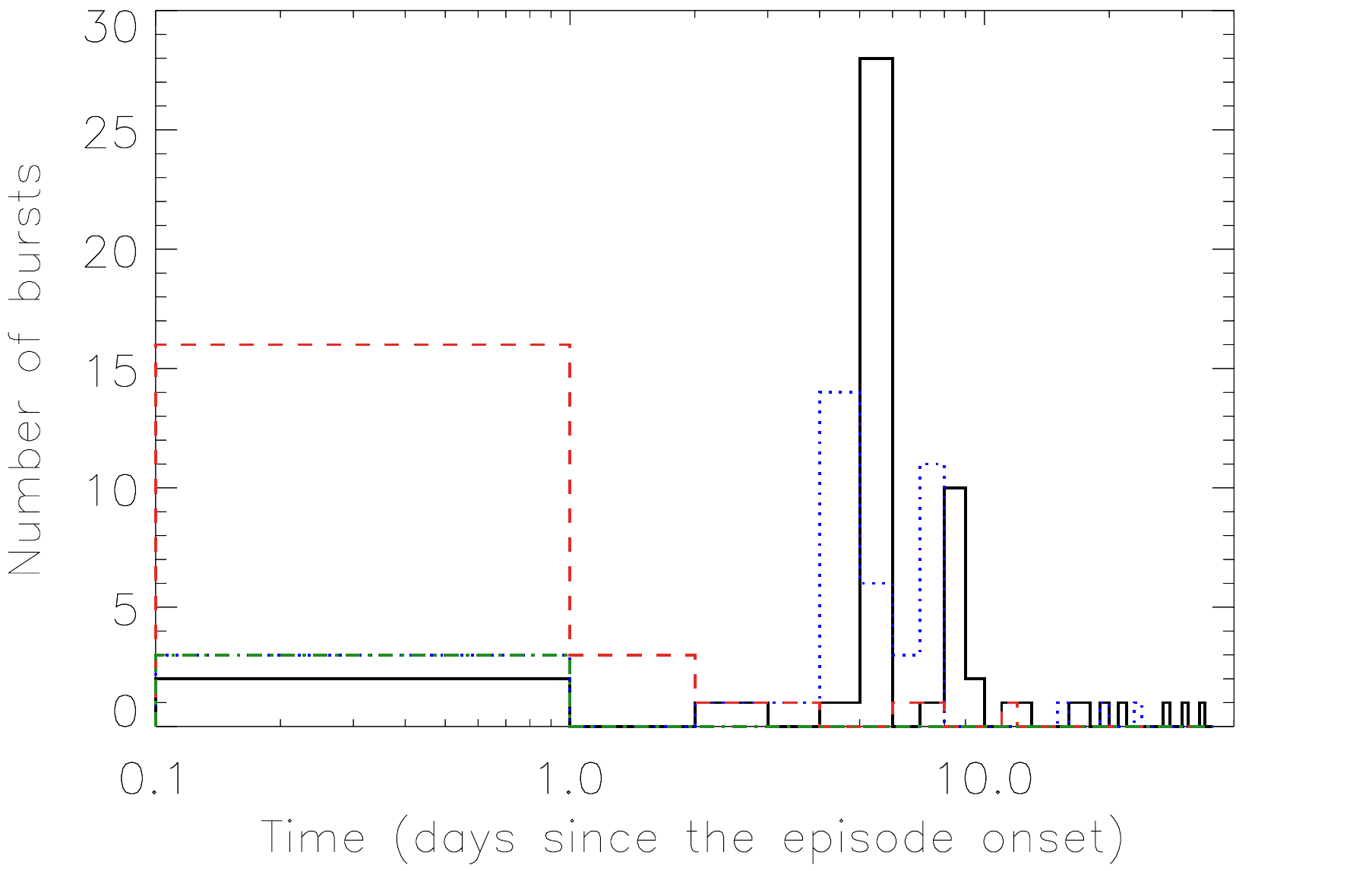}
\caption{The number of bursts in each active episode of \sgrnos. The blue dashed-dotted line, red dashed line, blue dotted line and the black solid line represent the first, second, third and fourth active burst episodes, respectively. \label{fig:evonb}}
\end{figure*}

\citet{younes2017} also found the increase in the average flux to be larger, decaying more rapidly in the 2016 outburst. However, the flux decay was much smoother in the first two episodes. Interestingly, we noticed that outbursts with different temporal profiles also exhibit diverse short burst history. As presented in Figure \ref{fig:evonb} and Figure 6 in \citet{younes2017}, all or the majority of bursts in the 2014 \& 2015 active episodes happened on the first day of the episode, before subsequently decaying over $\sim$100 days. However, two episodes in 2016 started with two or three bursts, with the largest number of bursts being emitted $4-10$ days later. The two outbursts in 2016 were brighter at the onset than those in 2014 \& 2015 and quickly decayed to the quiescent level after the bursts subsided. This connection between bursts and outbursts strongly indicates that the total energy released in short bursts accelerated the fading of the persistent outburst (at least one component of the persistent emission).

\subsection{Comparison of burst properties during the four active episodes}

A prolific transient is defined here as a magnetar emitting more than ten bursts during an active burst episode \citep{gogus2014}. Prior to \sgrnos, such transients included SGRs 1627$-$41 \citep{woods1999,esposito2008}, J0501+4516 \citep{rea2009,gogus2010,lin2011} and J1550$-$5418 \citep{israel2010,vdh2012,vonkienlin2012}, each with only one or two burst active episodes. We reported four episodes from \sgr in the first three years since its discovery, making it the most prolific magnetar transient to date. 

\begin{figure*}
\includegraphics*[viewport=75 165 600 495, scale=0.5]{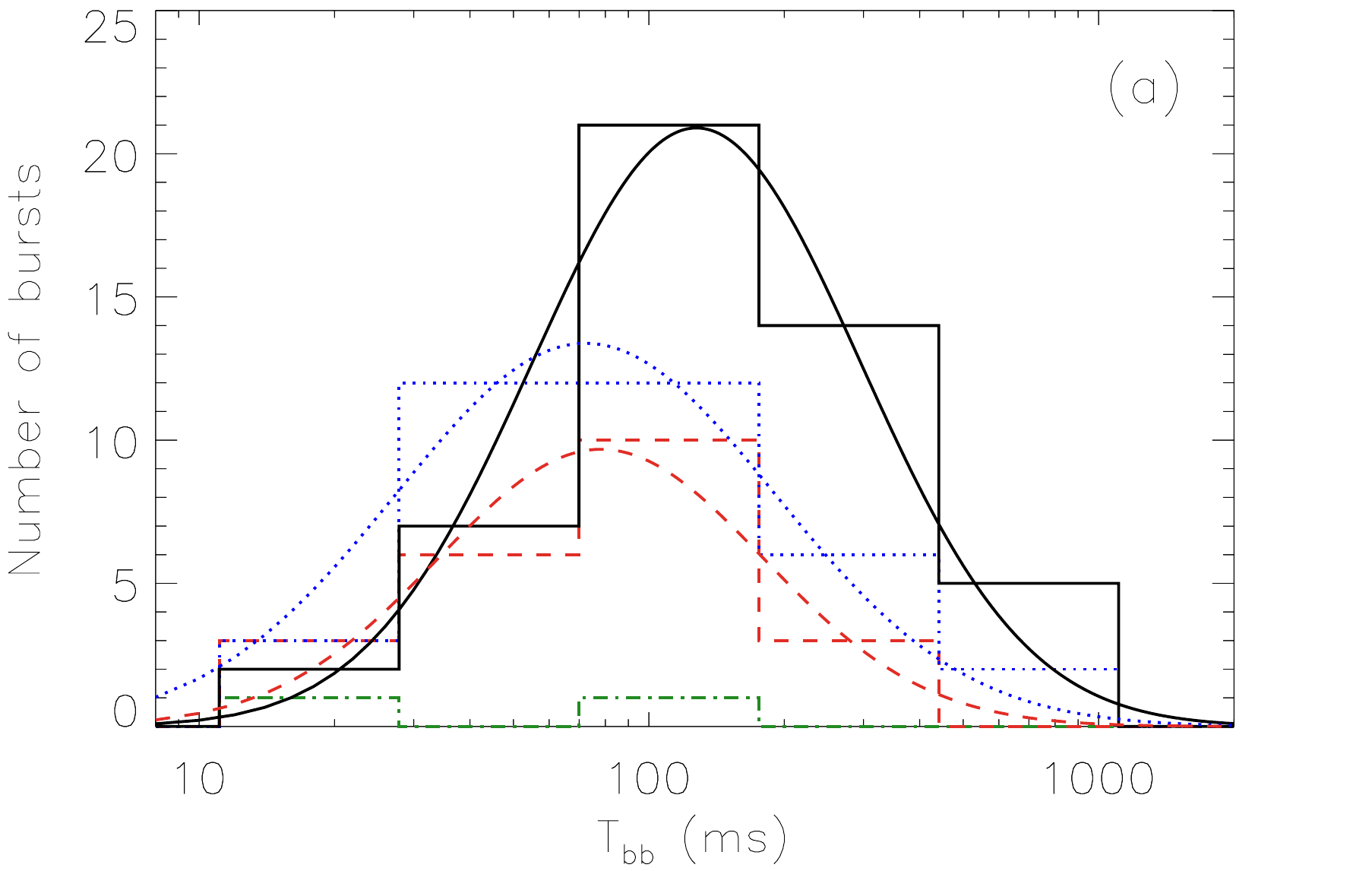}
\includegraphics*[viewport=75 165 600 495, scale=0.5]{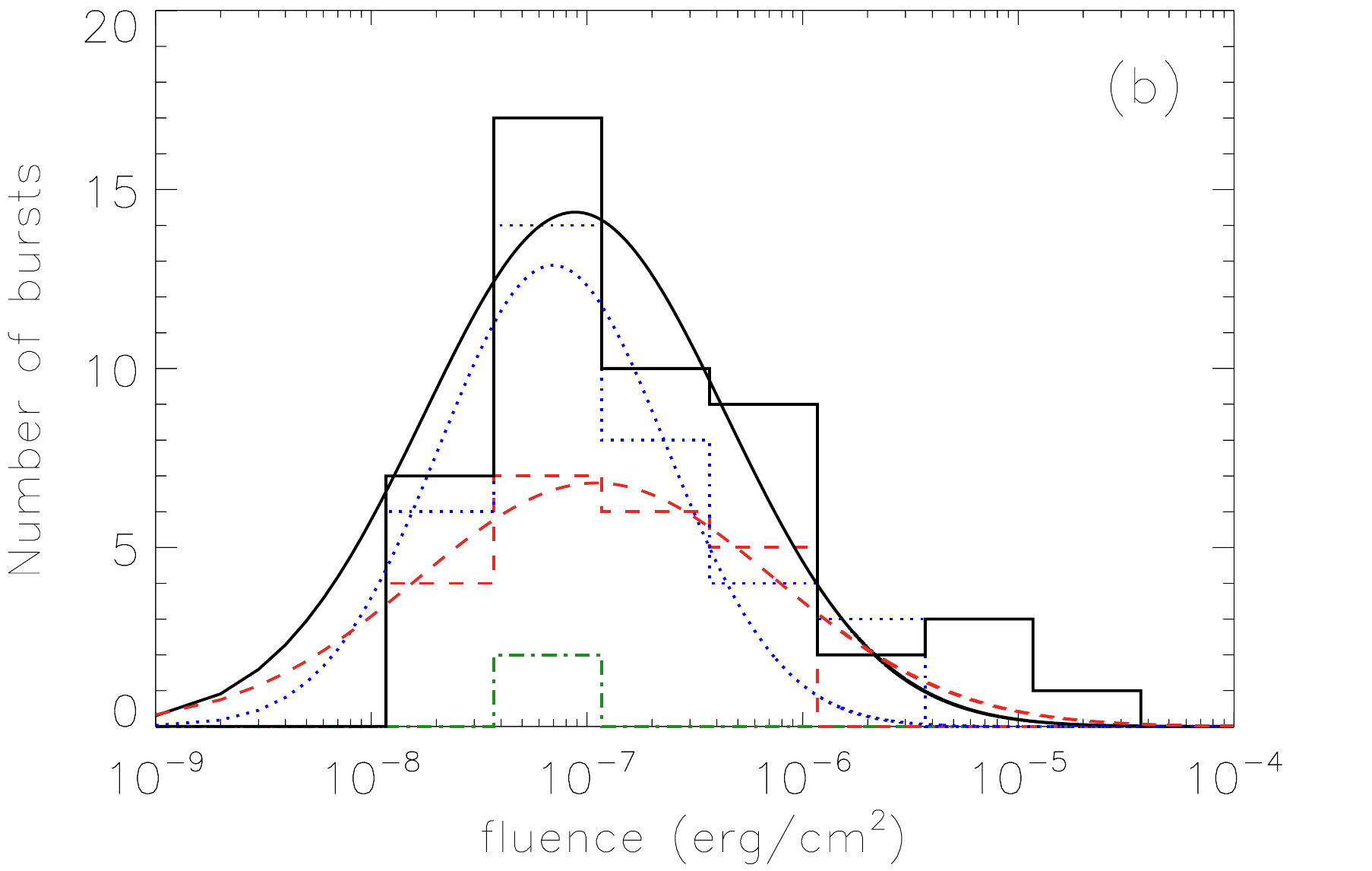}
\includegraphics*[viewport=75 165 600 495, scale=0.5]{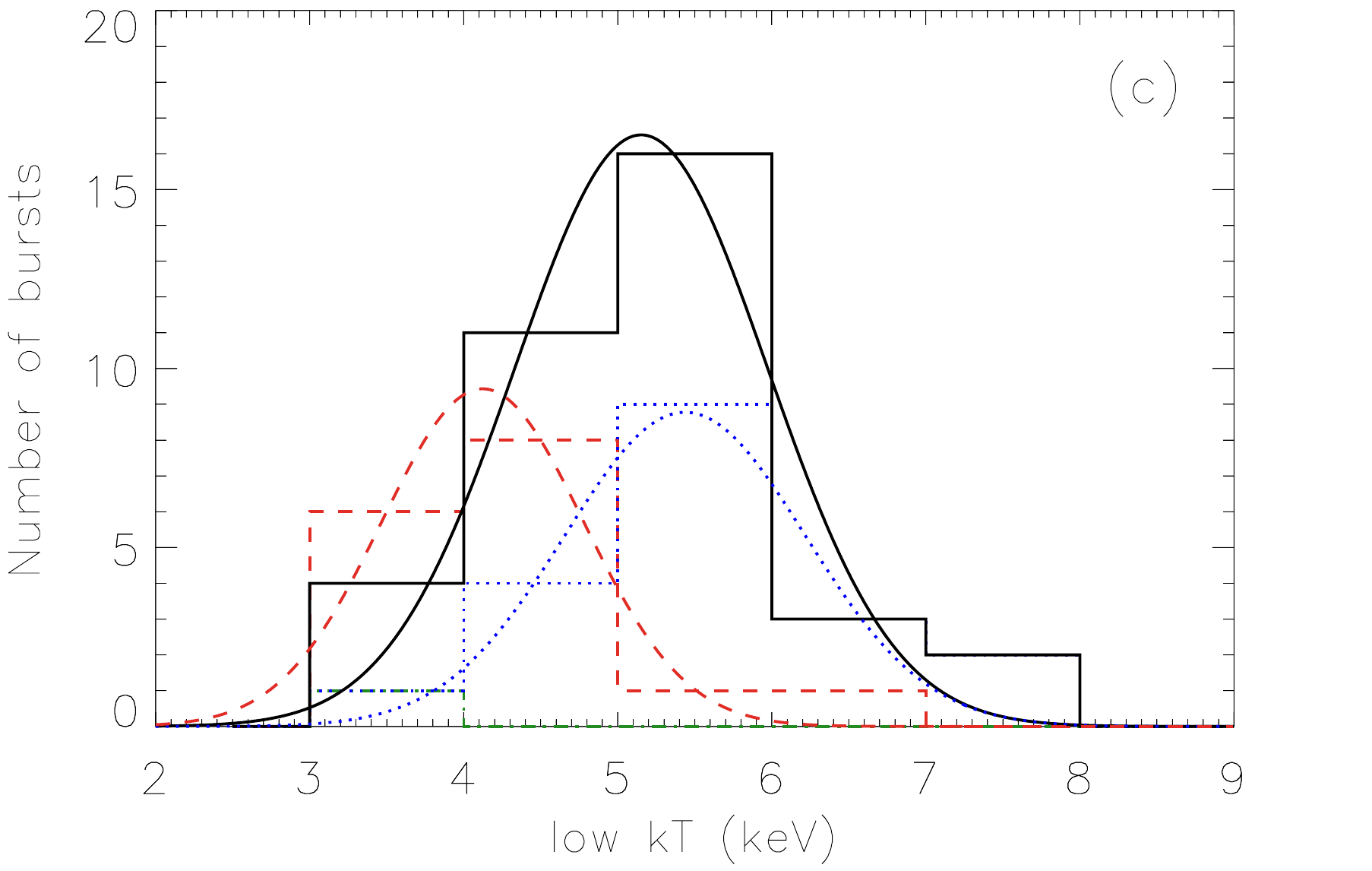}
\includegraphics*[viewport=75 165 600 495, scale=0.5]{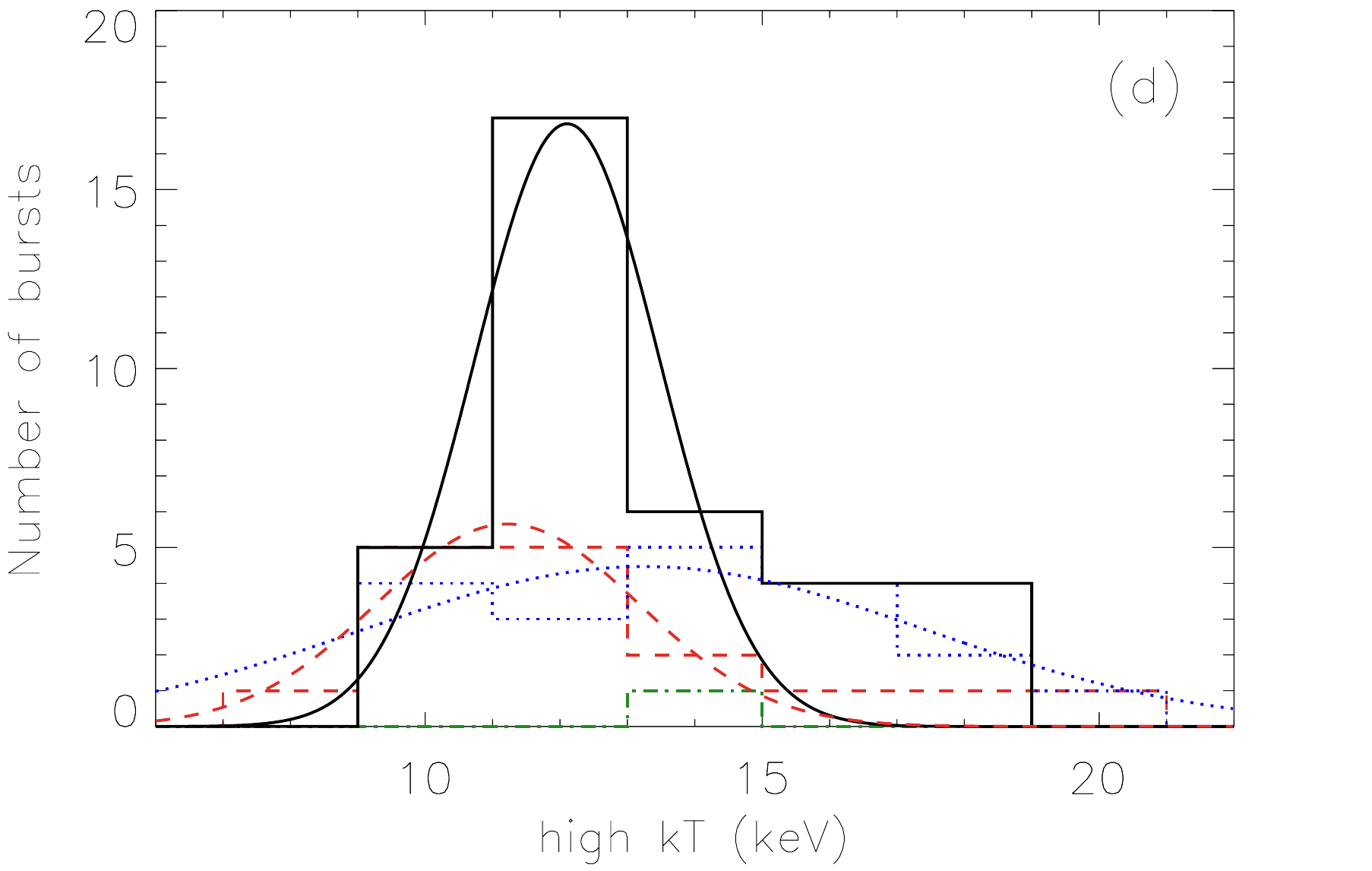}
\includegraphics*[viewport=75 165 600 495, scale=0.5]{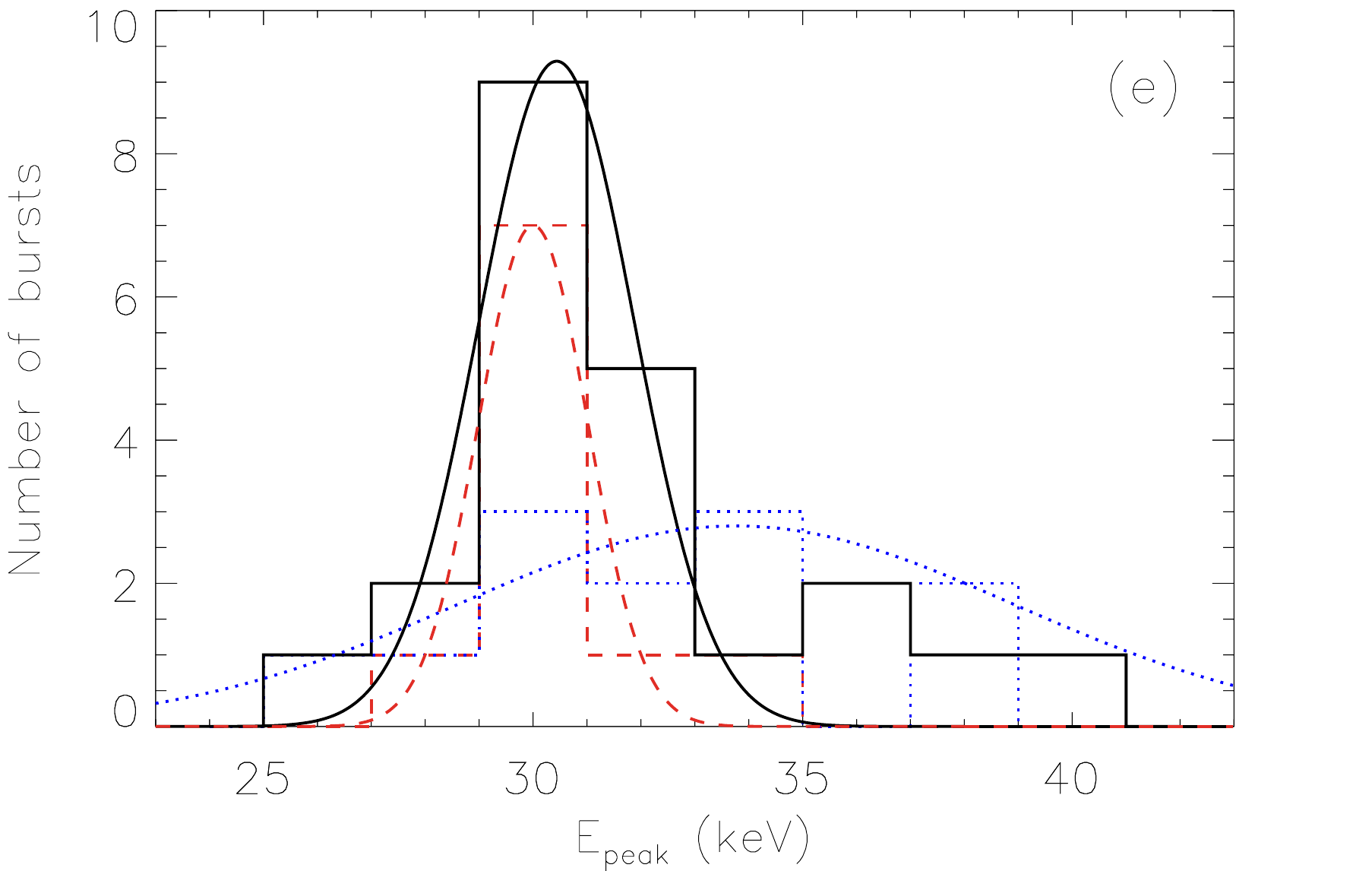}
\includegraphics*[viewport=75 165 600 495, scale=0.5]{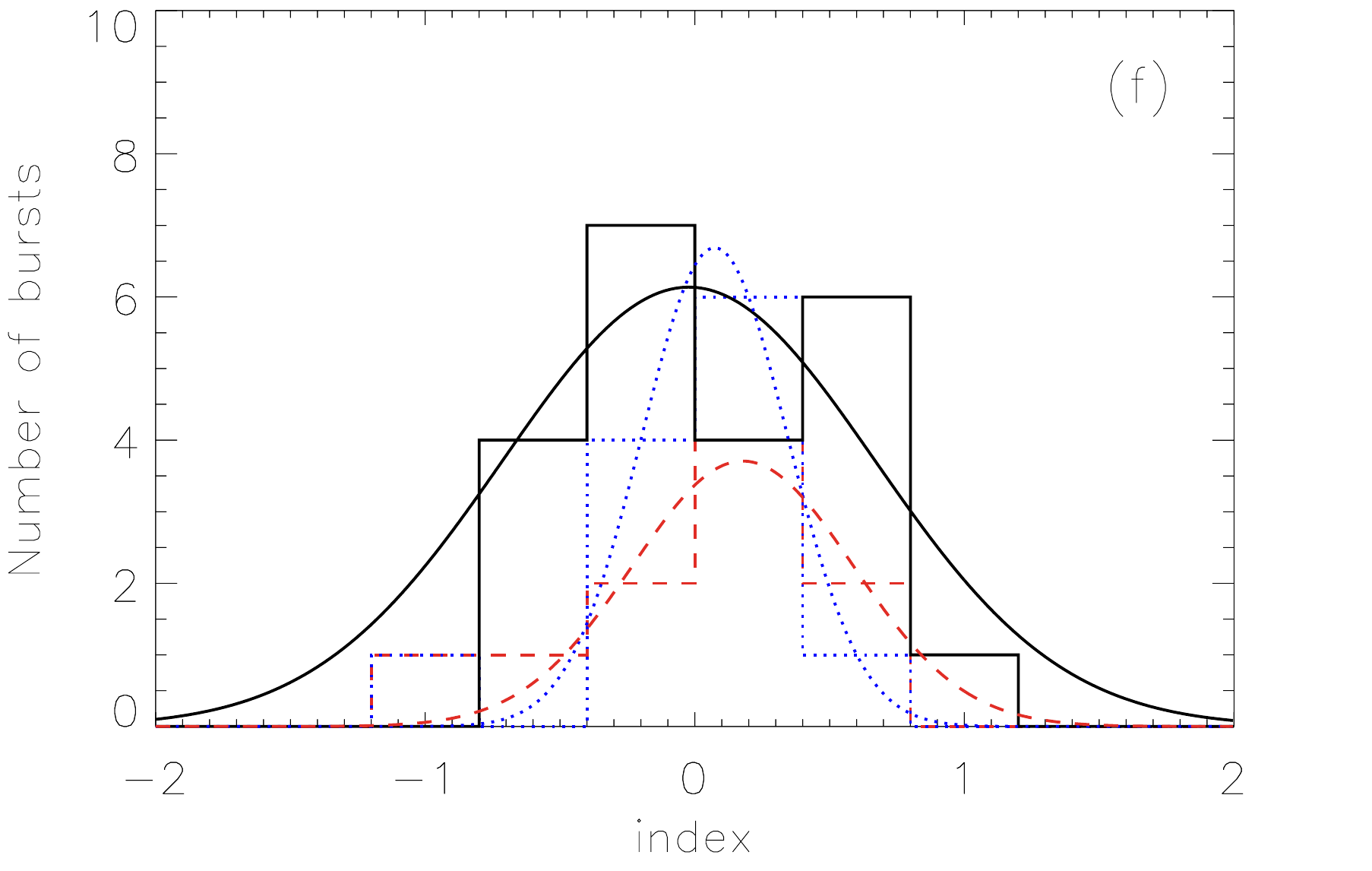}
\caption{Distributions of $T_{\rm bb}$ (a), fluence (b), along with the low (c) and high (d) temperature from fits using a BB+BB model fit are presented in the top and middle two panels, respectively. The $E_{\rm peak}$ (e) and index (f) distributions from fits to \sgr bursts using the COMPT model observed by GBM, a presented in the bottom two panels. The green dashed-dotted, red dashed, blue dotted and black solid histograms represent burst active episodes 1 through 4, respectively. The best Gaussian or log-Gaussian fits to the histograms are over-plotted in each panel.
\label{fig:evo}}
\end{figure*}

We also studied the distributions of temporal and spectral parameters of the bursts from each of the four active episodes (Figure \ref{fig:evo}). The mean values of the Gaussian or log-Gaussian fits to these distributions during the 2$^{nd}$, 3$^{rd}$ and 4$^{th}$ episodes, are listed in Table \ref{tab:evo}. The duration of bursts from the 2$^{nd}$ and 3$^{rd}$ episodes are shorter than the 4$^{th}$, while the average fluences are consistent (within error). Both BB temperatures from the BB+BB model are lower before 2016. Similar time evolution of burst characteristics was also reported for SGR J1550$-$5418 \citep{vonkienlin2012}.

\begin{deluxetable*}{ccccccc}
\tablenum{6}
\tablecaption{Mean parameters of GBM bursts during all active episodes. \label{tab:evo}}
\tablewidth{0pt}
\tablehead{
\colhead{Episode} & \colhead{$<T_{\rm{bb}}>$} & \colhead{$<Fluence>$} & \multicolumn{2}{c}{BB+BB} & \multicolumn{2}{c}{COMPT}  \\
 \colhead{} & \colhead{} & \colhead{} &\colhead{$<low~kT>$} &\colhead{$<high~kT>$} & \colhead{$<E_{\rm{peak}}>$} & \colhead{$<index>$} \\
\colhead{ } & \colhead{(ms) } &\colhead{ ($10^{-7} $erg cm$^{-2}$)} & \colhead{(keV)} & \colhead{(keV)} &\colhead{(keV)} &\colhead{} 
}
\startdata
2 & $78^{+17}_{-14}$ & $1.1^{+0.4}_{-0.3}$ & $4.1\pm0.1$ & $11.2\pm0.3$ & $30.0\pm0.4$ & $0.18\pm0.10$ \\
3 & $72^{+7}_{-6}$ & $0.7\pm0.2$ & $5.4\pm0.2$ & $13.3\pm0.9$ & $33.8\pm1.3$ & $0.07\pm0.05$\\
4 & $128^{+11}_{-10}$ & $0.9^{+0.3}_{-0.2}$ & $5.1\pm0.1$ & $12.1\pm0.4$ & $30.4\pm0.2$ & $-0.02\pm0.26$
\enddata
\end{deluxetable*}

\subsection{Burst properties and correlations}
In previous studies, the short burst duration distributions ($T_{90}$) followed a log-Gaussian shape, reaching a peak at $\sim 90-160~\rm{ms}$ \citep{gogus2001,gavriil2004,lin2011,vdh2012,collazzi2015}. The $T_{bb}$ of bursts from \sgr also followed a log-Gaussian distribution, with a mean value of $\sim94~\rm{ms}$. Although, $T_{bb}$ is typically longer than $T_{90}$, we find $<T_{bb}>$ for our sample to be at the lower end of known values. The mean $T_{90}$ from a log-Gaussian fit results in $\sim 75~\rm{ms}$, which is about two-thirds of the typical value for other magnetars.

\begin{figure*}
\includegraphics*[viewport=70 140 600 490, scale=0.9]{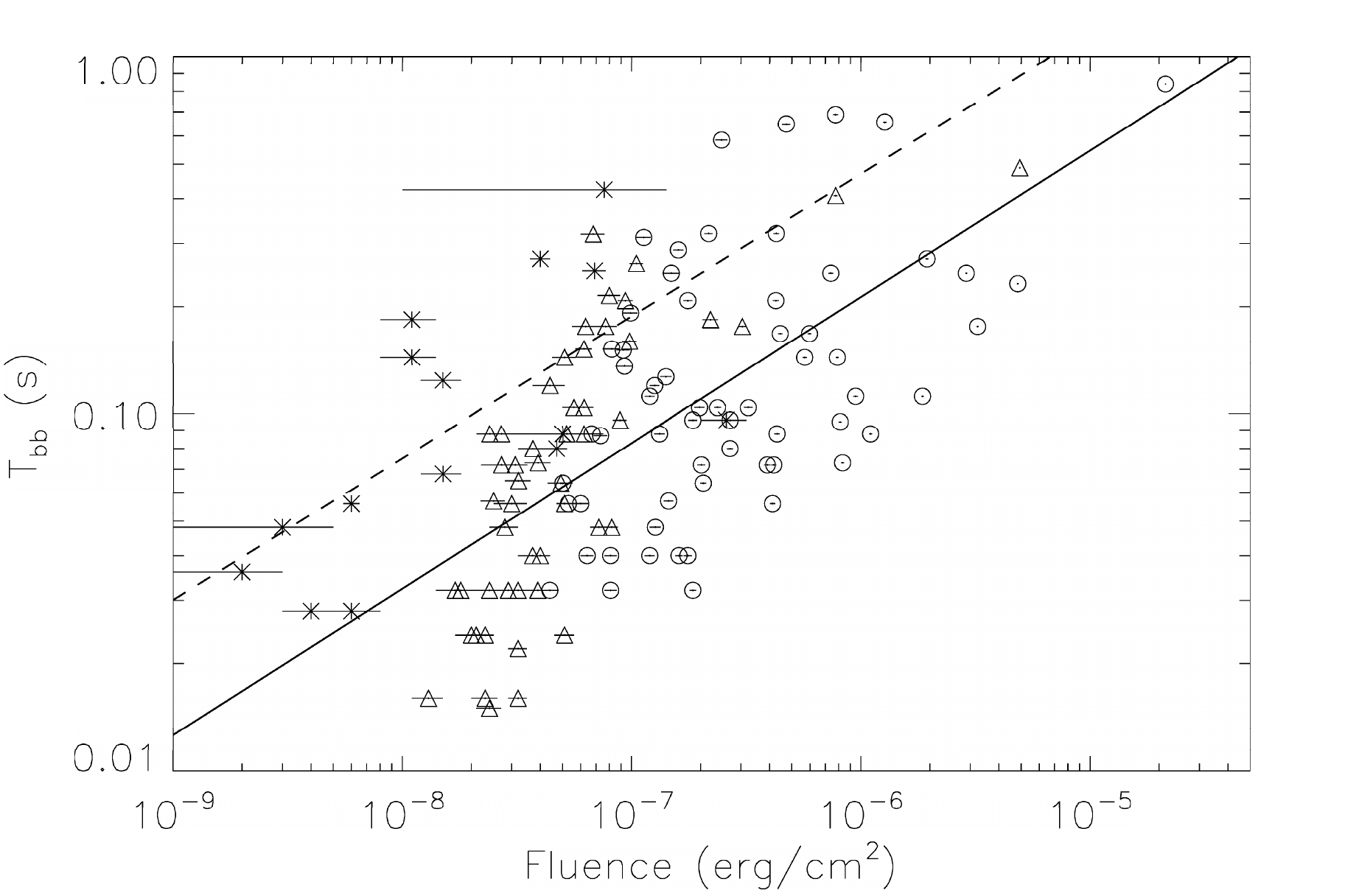}
\caption{The correlation between T$_{bb}$ and energy fluence. The stars, open circles and triangles represent BAT bursts, GBM triggered bursts and GBM un-triggered bursts, respectively. The dashed line is the best PL fit to 15 bursts only observed with the BAT. The solid line is the same fit to 112 GBM events. \label{fig:bstflnctbb}}
\end{figure*}

The correlation between the burst duration and fluence has been previously reported for several magnetars \citep{gogus2001,gavriil2004,vdh2012}. For bursts from \sgrnos, we also find that $T_{bb}$ is significantly correlated with energy fluence (Figure \ref{fig:bstflnctbb}). The Spearman rank correlation test coefficient is 0.6, with a chance probability of $1.2\times10^{-14}$ for 112 GBM bursts. We fit this correlation with a PL and find the index to be $0.41\pm0.04$ (using GBM bursts) and $0.40\pm0.13$ (using only BAT detected bursts), in very good agreement with each other. These values are also consistent with similar values for SGR 1806$-$20, SGR 1900+14 \citep{gogus2001} and SGR J1550$-$5418 \citep{vdh2012}, while being shallower than 1E 2259+586 \citep{gavriil2004}.

There are 80 bursts in our sample whose data is well fit using a BB+BB model. The temperatures of the two blackbody components are consistent with those reported for other magnetars (e.g., \citealt{lin2011, vdh2012, collazzi2015}). We do not find a correlation between the temperatures of the two BB components. \citet{kozlova2016} fit the time-integrated spectrum between $20-300~\rm{keV}$ for the intermediate flare from \sgr with the same model. They found a similar temperature for the hard BB component, and a higher temperature ($\sim6.4~\rm{keV}$) for the soft BB component. The difference may be due to the low-energy cutoff of the spectrum, which is 10~keV more than that of GBM. 

The relationship between the BB emission areas ($R^2$), and temperatures (see the lower right panel of Figure \ref{fig:BBBBcor}), indicates that the hotter BB components are from a smaller area. The Spearman test yields a correlation coefficient of -0.8 with a chance probability of $1.1\times10^{-20}$. We fit this correlation with a PL and obtain an index of $-10.6\pm1.8$. The correlation for the cooler BB components is not significant. The emission area of the cool component spreads around a mean value of $\sim161~\rm{km}^2$. If we describe both BB components with one PL, the index for the best fit is $-4.2\pm0.3$, which is consistent with the index of a BB with steady luminosity. These results agree with those for other magnetars, e.g., SGR J1550$-$5418 \citep{vdh2012} and SGR J0501+4516 \citep{lin2011}. The evolution of the emission area with BB temperature, the strong correlation between the emission areas and luminosities of both BB components (Figure \ref{fig:BBBBcor}), as well as the equally divided energy for both BB fits, possibly indicates that the two BB components are strongly connected events despite emanating from very different regions. As discussed in \citet{vdh2012,younes2014}, a detailed modeling of the dynamic fireball in the magnetosphere is required to understand the actual physical process.

The $E_{peak}$ and PL index of the COMPT model are quite different for \sgr when compared with other magnetars. The distribution of $E_{peak}$ values for all bursts peaks at $\sim 30~\rm{keV}$, which is softer than SGR\,J1550$-$5418 \citep{vdh2012} and SGR\,J0501+4516 \citep{lin2011} ($\sim 40 \rm{keV}$). In the Comptonization process, the photons can be scattered up to a higher energy consistent with the electron temperature. The softer $E_{peak}$ indicates that either the temperature of the plasma covering \sgr is slightly cooler than other magnetars or, the \sgr bursts are emitted from a relatively cooler region of \sgrnos. 

The average spectral index is $\sim -0.1$, similar to that of bursts from SGR\,J0501+4516 \citep{lin2011}, while harder than the index measured in SGR\,J1550$-$5418 bursts ($\sim -1$) \citep{vdh2012}. As discussed in \citet{lin2011}, the index of Compton up-scattering depends on the mean energy change per collision and the mean amount of scattering. However, this parameter can change in magnetars due to the presence of a strong magnetic field. The evolution of index with burst fluence is also very interesting. The dimmer bursts present the flattest spectrum that can be produced with Compton up-scattering, while the brighter bursts exhibit a more thermalized environment. A similar correlation was only reported by \citet{younes2014}, who studied the time-resolved spectra of SGR\,J$1550-5418$. These results stress that time-resolved analysis is crucial in understanding how the emission properties evolve through a magnetar burst. We will present the results of the time-resolved analysis for \sgr bursts in an upcoming paper.

\clearpage
\begin{longrotatetable}
\begin{deluxetable*}{cccccccccc}
\tablecolumns{10}
\tablenum{4}
\tablecaption{The Duration, Localization and Fluence for each \sgr burst. \label{tab:burstlist}}
\tablewidth{0pt}
\tablehead{
\colhead{ID} & \colhead{Instrument} & \colhead{$t_{burst}$} & \colhead{$T_{BB}$} & \colhead{$T_{90}$} & \colhead{$T_{50}$} & \colhead{ra} & \colhead{dec} & \colhead{err} & \colhead{Fluence}  \\
\colhead{} & \colhead{} & \colhead{in UTC} & \colhead{(s)} & \colhead{(s)} & \colhead{(s)} & \colhead{($^{\circ}$)} & \colhead{($^{\circ}$)} & \colhead{($^{\circ}$)} & \colhead{($10^{-7}~\rm{erg}~\rm{cm}^{-2}$)}  
}
\startdata
1	 & 	GBM$^{U}$	 & 	2014-07-05 09:32:48.643	 & 	0.088	 & 	$	0.112\pm0.051	$	 & 	$	0.048\pm0.011	$	 & 	297.28 	 & 	18.18 	 & 	6.57 	 & 	$	0.62 	\pm	0.05 	$	\\
	 & 	BAT$^{T}$	 & 		 & 	0.088	 & 		\nodata		 & 		\nodata		 & 	\nodata	 & 	\nodata	 & 	\nodata	 & 	$	0.41 	\pm	0.26 	$	\\
2	 & 	GBM$^{U}$	 & 	2014-07-05 09:37:34.484	 & 	0.022	 & 	$	0.024\pm0.040	$	 & 	$	0.008\pm0.011	$	 & 	291.63 	 & 	20.72 	 & 	9.65 	 & 	$	0.32 	\pm	0.03 	$	\\
	 & 	BAT$^{U}$	 & 		 & 	0.100	 & 		\nodata		 & 		\nodata		 & 	\nodata	 & 	\nodata	 & 	\nodata	 & 	$	0.30 	\pm	0.03 	$	\\
3	 & 	BAT$^{U}$	 & 	2014-07-05 09:41:06.139	 & 	0.124	 & 		\nodata		 & 		\nodata		 & 	\nodata	 & 	\nodata	 & 	\nodata	 & 	$	0.15 	\pm	0.03 	$	\\
4	 & 	BAT$^{T}$	 & 	2015-02-22 12:16:26.304	 & 	0.088	 & 		\nodata		 & 		\nodata		 & 	\nodata	 & 	\nodata	 & 	\nodata	 & 	$	0.50 	\pm	0.25 	$	\\
5	 & 	BAT$^{T}$	 & 	2015-02-22 12:31:11.348	 & 	0.080	 & 		\nodata		 & 		\nodata		 & 	\nodata	 & 	\nodata	 & 	\nodata	 & 	$	0.47 	\pm	0.05 	$	\\
6	 & 	GBM$^{U}$	 & 	2015-02-22 07:16:56.058	 & 	0.016	 & 	$	0.008\pm0.029	$	 & 	$	0.000\pm0.011	$	 & 	305.65 	 & 	15.84 	 & 	11.93 	 & 	$	0.13 	\pm	0.02 	$	\\
\enddata
\end{deluxetable*}
\end{longrotatetable}

\clearpage
\begin{longrotatetable}
\begin{deluxetable*}{cccccccccc}
\tablecolumns{10}
\tablenum{5}
\tablecaption{Spectral fit parameters for all \sgr bursts. \label{tab:burstspec}}
\tablewidth{0pt}
\tablehead{
\colhead{ID} & \colhead{Instrument} & \multicolumn{3}{c}{BB+BB or BB} &  \multicolumn{3}{c}{COMPT or PL}  &  \multicolumn{2}{c}{OTTB}\\
\colhead{} & \colhead{} & \colhead{kT$_1$} & \colhead{kT$_2$} & \colhead{$\chi^2/d.o.f$} & \colhead{$\Gamma$} & \colhead{kT} & \colhead{$\chi^2/d.o.f$}   &\colhead{kT} & \colhead{$\chi^2/d.o.f$}\\
\colhead{} & \colhead{} & \colhead{(keV)} & \colhead{(keV)} & \colhead{} & \colhead{} & \colhead{(keV)} & \colhead{}   &\colhead{(keV)} & \colhead{}
}
\startdata
1	 & 	GBM$^{U}$	 & 	$	3.31 	_{	-0.60 	}^{+	0.58 	}	$	 & 	$	13.38 	_{	-1.80 	}^{+	2.01 	}	$	 & 	$	316.69 	/	265 	$	 & 	$	-2.35 	_{	-0.11 	}^{+	0.10 	}	$	 & 		\nodata							 & 	$	318.32 	/	267 	$	 & 	$	28.70 	_{	-3.48 	}^{+	4.04 	}	$	 & 	$	320.56 	/	267 	$	\\
	 & 	BAT$^{T}$	 & 	$	4.57 	_{	-0.76 	}^{+	0.85 	}	$	 & 	$	14.63 	_{	-2.25 	}^{+	3.18 	}	$	 & 	$	7.07 	/	9 	$	 & 	$	-2.46 	_{	-0.15 	}^{+	0.14 	}	$	 & 		\nodata							 & 	$	9.11 	/	11 	$	 & 	$	28.43 	_{	-3.81 	}^{+	4.45 	}	$	 & 	$	10.15 	/	11 	$	\\
2	 & 	GBM$^{U}$	 & 	$	7.45 	_{	-0.57 	}^{+	0.57 	}	$	 & 									 & 	$	283.25 	/	266 	$	 & 	$	-2.37 	_{	-0.13 	}^{+	0.12 	}	$	 & 		\nodata							 & 	$	274.28 	/	266	$	 & 	$	25.06 	_{	-3.19 	}^{+	3.80 	}	$	 & 	$	268.61 	/	266 	$	\\
	 & 	BAT$^{T}$	 & 	$	4.07 	_{	-1.18 	}^{+	1.32 	}	$	 & 	$	13.17 	_{	-2.13 	}^{+	3.12 	}	$	 & 	$	2.39 	/	7 	$	 & 	$	-2.40 	_{	-0.16 	}^{+	0.15 	}	$	 & 		\nodata							 & 	$	4.21 	/	9 	$	 & 	$	30.15 	_{	-4.36 	}^{+	5.23 	}	$	 & 	$	3.51 	/	9 	$	\\
3	 & 	BAT$^{T}$	 & 				\nodata					 & 				\nodata					 & 		\nodata				 & 	$	-1.56 	_{	-0.35 	}^{+	0.35 	}	$	 & 		\nodata							 & 	$	4.09 	/	2 	$	 & 		\nodata							 & 		\nodata				\\
4	 & 	BAT$^{T}$	 & 	$	3.05 	_{	-0.84 	}^{+	0.92 	}	$	 & 	$	10.07 	_{	-1.94 	}^{+	2.30 	}	$	 & 	$	11.43 	/	9 	$	 & 	$	-3.29 	_{	-0.37 	}^{+	0.31 	}	$	 & 		\nodata							 & 	$	14.53 	/	11 	$	 & 	$	15.78 	_{	-3.56 	}^{+	4.21 	}	$	 & 	$	14.03 	/	11 	$	\\
5	 & 	BAT$^{T}$	 & 	$	3.56 	_{	-0.91 	}^{+	1.11 	}	$	 & 	$	10.00 	_{	-1.13 	}^{+	1.92 	}	$	 & 	$	14.27 	/	13 	$	 & 	$	-2.81 	_{	-0.11 	}^{+	0.11 	}	$	 & 		\nodata							 & 	$	20.84 	/	17 	$	 & 	$	21.23 	_{	-1.79 	}^{+	1.95 	}	$	 & 	$	14.80 	/	15 	$	\\
6	 & 	GBM$^{U}$	 & 	$	5.70 	_{	-0.65 	}^{+	0.73 	}	$	 & 				\nodata					 & 	$	173.25 	/	197 	$	 & 	$	-2.48 	_{	-0.24 	}^{+	0.22 	}	$	 & 		\nodata							 & 	$	178.00 	/	197 	$	 & 		\nodata							 & 		\nodata				\\
\enddata

\end{deluxetable*}
\end{longrotatetable}

\acknowledgments
L.L. acknowledges support from the National Natural Science Foundation of China (grant no. 11703002 and 11543004) and the Fundamental Research Funds for the Central Universities. Y.K acknowledges the support from the Scientific and Technological Research Council of Turkey (T\"{U}B\.{I}TAK grant no. 118F344). 

\software{HEAsoft (v624; HEASARC 2014), 
XSPEC (Arnaud 1996),
RMFIT (v40rc1; Mallozzi 2000, Preece 2011),
GBMDRM (v2.0)}

\appendix

\section{Exploring the GBM location accuracy with \sgr bursts}

\begin{figure*}[h!]
\includegraphics*[viewport=70 140 600 490, scale=0.8]{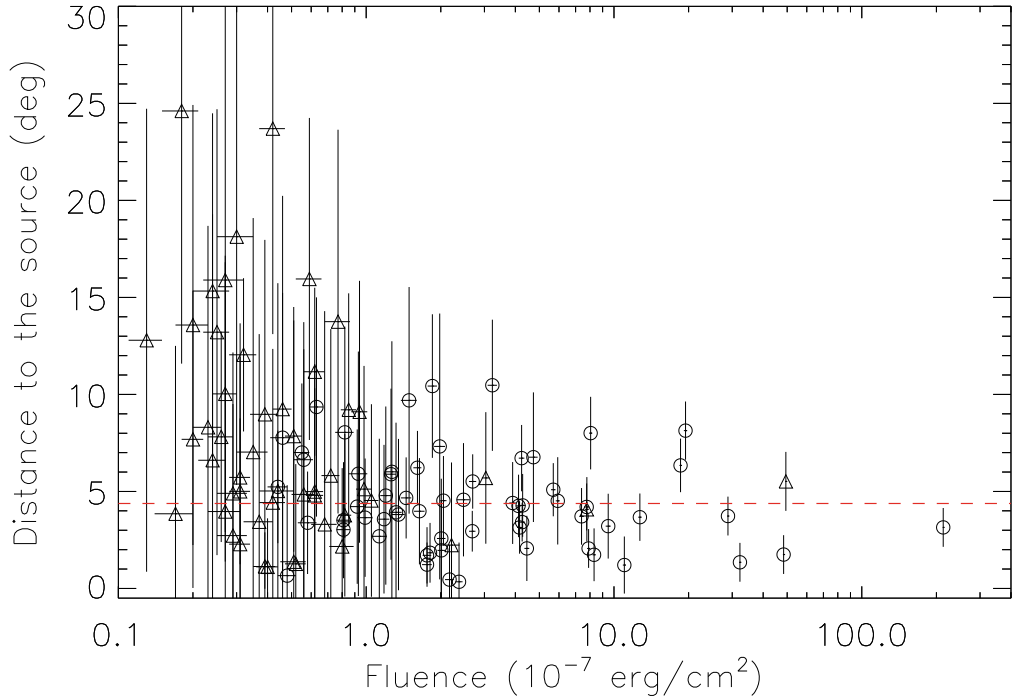}
\caption{The offset of the measured locations to each burst relative to the \sgr position as a function of the burst fluence. The open circles and triangles represent the triggered and un-triggered bursts, respectively. The red dashed line is the weighted average of the offset. \label{fig:bstlocsyserr}}
\end{figure*}

The soft spectrum and large number of repeated events make short bursts from magnetars a good database to study the systematic location accuracy of GBM in the soft energy band ($<50~\rm{keV}$). We plot the offset to the \sgr position \textit{v.s.} the burst fluence in Figure \ref{fig:bstlocsyserr}. Weaker bursts present larger offsets as well as larger statistical uncertainties. Brighter bursts with much smaller statistical uncertainties also show offsets with several degrees, which come from systematical effects. This is also pointed out by a similar study using 21 bursts triggered GBM from SGR~J1550$-$5418 \citep{vonkienlin2012}. We set the reverse of the statistical error as the weight of each burst and calculate the weighted mean of the distance to the known source position as $\sim4.4^{\circ}$. This systematic offset for the soft band of GBM is close to the result of $3.7^{\circ}$ from \citet{connaughton2015} for Gamma-Ray Bursts in the harder GBM energy band (50-300 keV). We note here, however, that a detailed study of the systematic uncertainties of the GBM location in the soft band should include short bursts from other magnetars as well as a more careful modeling, both of which are not within the scope of this paper.

{}

\end{document}